\documentclass[prx,aps,twocolumn,superscriptaddress]{revtex4-2}

\usepackage{graphicx}
\usepackage{amssymb,amsfonts,amsmath}
\usepackage{color}
\usepackage{ulem}
\usepackage[hidelinks]{hyperref}
\usepackage{bbm}
\usepackage{bbold}
\usepackage[utf8]{inputenc}

\usepackage{tikz}
\usetikzlibrary{shapes}
\usetikzlibrary{patterns}
\usetikzlibrary{angles, quotes}

\newcommand{\rv}{{\mathbf r}}
\newcommand{\Rv}{{\mathbf R}}

\newcommand{\Tr}{{\rm Tr}\,}
\newcommand{\e}{{\rm e}}

\newcommand{\Jv}{{\bf J}}

\newcommand{\pv}{{\bf p}}

\newcommand{\Pv}{{\bf P}}

\newcommand{\Fv}{{\bf F}}
\newcommand{\fv}{{\bf f}}

\newcommand{\vel}{{\bf v}}

\newcommand{\msphantom}[1]{$\ldots$}

\newcommand{\eqr}[1]{Eq.~\eqref{#1}}

\newcommand{\unity}{{\mathbbm 1}}

\newcommand{\mydelete}[1]{{}}
\newcommand{\taub}{{\boldsymbol\tau}}

\newcommand{\rmint}{{\rm int}}

\newcommand{\rmext}{{\rm ext}}

\newcommand{\rmid}{{\rm id}}
\newcommand{\rmacc}{{\rm acc}}
\newcommand{\bsig}{{\boldsymbol\sigma}}
\newcommand{\Sv}{{\bf S}}

\DeclareMathOperator{\cov}{cov}

\newcommand{\calU}{\,{\cal U}}
\newcommand{\mystar}{{\bullet}}
\newcommand{\Bare}{{\bf S}}
\newcommand{\Cv}{{\bf C}}
\newcommand{\size}{a}
\newcommand{\timescale}{t_{\rm MD}}

\begin{document}

\title{Dynamical gauge invariance of statistical mechanics}

\author{Johanna M\"uller}
\affiliation{Theoretische Physik II, Physikalisches Institut, 
  Universit{\"a}t Bayreuth, D-95447 Bayreuth, Germany}

\author{Florian Samm\"uller}
 \affiliation{Theoretische Physik II, Physikalisches Institut, 
   Universit{\"a}t Bayreuth, D-95447 Bayreuth, Germany}

\author{Matthias Schmidt}
\affiliation{Theoretische Physik II, Physikalisches Institut, 
  Universit{\"a}t Bayreuth, D-95447 Bayreuth, Germany}
\email{Matthias.Schmidt@uni-bayreuth.de}

\date{24 April 2025}

\begin{abstract}
We investigate gauge invariance against phase space shifting in
nonequilibrium systems, as represented by time-dependent many-body
Hamiltonians that drive an initial ensemble out of thermal
equilibrium.  The theory gives rise to gauge correlation functions
that characterize spatial and temporal inhomogeneity with microscopic
resolution on the one-body level. Analyzing the dynamical gauge
invariance allows one to identify a specific localized shift gauge
current as a fundamental nonequilibrium observable that characterizes
particle-based dynamics. When averaged over the nonequilibrium
ensemble, the shift current vanishes identically, which constitutes an
exact nonequilibrium conservation law that generalizes the
Yvon-Born-Green equilibrium balance of the vanishing sum of ideal,
interparticle, and external forces. Any given observable is associated
with a corresponding dynamical hyperforce density and hypercurrent
correlation function. An exact nonequilibrium sum rule interrelates
these one-body functions, in generalization of the recent hyperforce
balance for equilibrium systems. We demonstrate the physical
consequences of the dynamical gauge invariance using both harmonically
confined ideal gas setups, for which we present analytical solutions,
and molecular dynamics simulations of interacting systems, for which
we demonstrate the shift current and hypercurrent correlation
functions to be accessible both via finite-difference methods and via
trajectory-based automatic differentiation. We show that the theory
constitutes a starting point for developing nonequilibrium
reduced-variance sampling algorithms and for investigating
thermally-activated barrier crossing.
\end{abstract}

\maketitle

\section{Introduction}
\label{SECintroduction}

Gauge invariance is one of the arguably most powerful construction
principles of theoretical physics, with profound consequences that
range from classical electrodynamics to the standard model of particle
physics and beyond. The gauge freedom in a given theory is intimately
linked to the validity of exact relationships, which typically possess
the form of conservation laws, such as those for electrical and more
general charges in the above fundamental theories. The link between
the gauge freedom and the associated conservation laws is provided by
Noether's theorem of invariant variations \cite{noether1918,
  byers1998, brading2002, read2022book}.  In a typical setting, one
applies the theorem to an action functional, which serves the dual
purpose of generating the dynamics from an associated variational
principle and providing the conservation laws via the gauge freedom.

For the deterministic dynamics of particle-based systems a range of
variational approaches exists, perhaps most notably in the form of
Hamilton's principle of stationary action~\cite{goldstein2002}. A
statistical mechanical description of the dynamics is then typically
based on the Liouville equation for the time evolution of the
many-body probability distribution on phase space \cite{zwanzig2001,
  hansen2013}. While direct numerical solution of the Liouville
equation for realistic systems comes at prohibitive cost,
trajectory-based molecular dynamics simulations provide a powerful
alternative \cite{hansen2013, frenkel2023book, schmid2022editorial}.
Corresponding theoretical developments in nonequilibrium statistical
mechanics \cite{zwanzig2001, hansen2013} are based on a range of
techniques and coarse-graining strategies~\cite{schilling2022},
including mode-coupling theory of the glass
transition~\cite{goetze2009, janssen2018}, stochastic thermodynamics
and its associated fluctuation theorems~\cite{seifert2012}, as well as
dynamical density \cite{evans1979, archer2004, tevrugt2020review,
  tevrugt2022perspective} and power functional theory
\cite{schmidt2022rmp, schmidt2018md, delasheras2023perspective,
  zimmermann2024ml}.

The dynamical behaviour and nonequilibrium phenomena that occur in
soft matter systems encompass a broad spectrum of physical systems and
effects \cite{hansen2013, evans2019physicsToday}.  Representative
topical examples include the emergence of solitons under overdamped
Brownian dynamics \cite{antonov2022}, the physics of dynamical
exclusion processes \cite{lips2018}, the dynamics of confined
electrolytes \cite{pireddu2024}, the non-trivial properties of the
electrical noise at the nanoscale \cite{minh2023jcpConfined,
  marbach2021reservoirs}, the aging of glasses \cite{janzen2023}, and
features of memory in generalised Langevin dynamics
\cite{winter2023ml}.

Owing to the significant increase of the degree of complexity of the
nonequilibrium problem over the equilibrium physics, the body of
available {\it exact} dynamical results is arguably less well
developed than in equilibrium, where a breadth of statistical
mechanical sum rules is available \cite{hansen2013, evans1979,
  baus1984, baus1992, evans1990, henderson1992, triezenberg1972}.
Notable fundamental relations that are relevant in nonequilibrium
include the Yvon theorem~\cite{yvon1935, goetze2009, hansen2013}, as
is based on partial integration on phase space, and Hirschfelder's
hypervirial theorem~\cite{hirschfelder1960} which generalizes the
standard virial theorm \cite{hansen2013} to arbitrary observables.
Furthermore, stochastic thermodynamics provides a significant body of
fluctuation theorems and further results \cite{seifert2012}.

Noether's theorem \cite{noether1918, byers1998, brading2002,
  read2022book} has only recently been applied to statistical physics
as a systematic construction principle within a variety of theoretical
approaches \cite{baez2013markov, marvian2014quantum, sasa2016,
  sasa2019, revzen1970, budkov2022, bravetti2023, brandyshev2023,
  beyen2024, beyen2024generic}. In particular, based on a specific
`shifting' operation on phase space, a broad range of equilibrium
statistical mechanical sum rules could both be reproduced and extended
\cite{hermann2021noether, hermann2022topicalReview,
  hermann2022variance, hermann2022quantum, sammueller2023whatIsLiquid,
  hermann2023whatIsLiquid, tschopp2022forceDFT, robitschko2024any,
  mueller2024gauge, mueller2024whygauge}. The underlying localized
shifting operation on phase space was identified subsequently as a
gauge transformation of equilibrium statistical mechanics
\cite{mueller2024gauge, mueller2024whygauge}; see
Refs.~\cite{rotenberg2024spotted, miller2025physicsToday} for recent
popular accounts.  In equilibrium, making use of the nontrivial Lie
algebra structure gives a promising perspective on the future
development of novel equilibrium sampling methods and exact sum rule
construction \cite{mueller2024gauge, mueller2024whygauge}.

\begin{figure}[!t]
  \vspace{1mm}
  \includegraphics[page=1,width=.99\columnwidth]{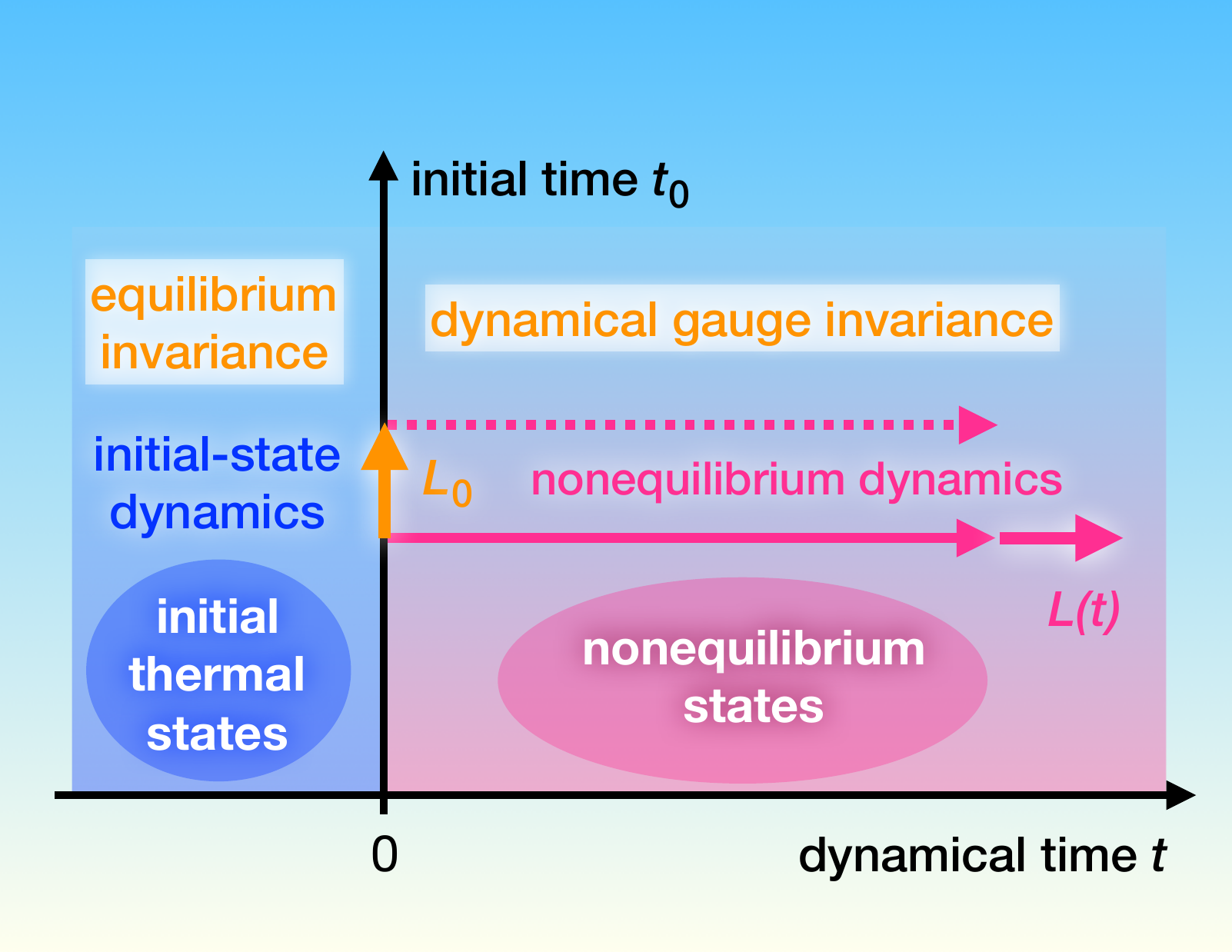}
  \caption{Illustration of the temporal structure of the
    nonequilibrium gauge theory. The initial state is in thermal
    equilibrium at inverse temperature~$\beta$ up to time $0$, with
    the dynamics being governed by the stationary Liouvillian~$L_0$
    (orange vertical arrow).  The time-dependent Liouvillian~$L(t)$
    then creates nonequilibrium dynamics (magenta horizontal arrows)
    at times $t\geq 0$.}
\label{FIG1}
\end{figure}

Here we investigate the consequences of the statistical mechanical
gauge invariance for the dynamics of particle-based systems in general
{\it nonequilibrium} setups, as described by the Liouville
equation. We introduce dynamical versions of the static differential
shifting operators \cite{mueller2024gauge, mueller2024whygauge}. These
allow one, upon systematic analysis of the effects of dynamic shifting
on the temporal propagation of dynamical averages, to derive and to
validate very general exact nonequilibrium sum rules.  These
identities interrelate specific gauge correlation functions that
possess microscopically sharp dependence on position and on time.  In
particular the shift current, the dynamical hyperforce density, and
the hypercurrent correlation function emerge naturally in the theory
and they are interrelated by exact nonequilibrium sum rules. The limit
of equilibrium dynamics constitutes a special, yet nontrivial case, as
we demonstrate. A central mechanism of the framework is a specific
differential operation induced by the time evolution of the initial
equilibrium ensemble, see Fig.~\ref{FIG1} for an illustration.

As a simple yet useful and analytically tractable toy model, we apply
the dynamical gauge framework to the harmonically confined ideal gas
where switching at an initial time generates a nonequilibrium
situation. Analyzing the gauge correlation functions provides deep
insight into the motion of the ensemble. Even when reduced to
equilibrium, the theory yields nontrivial insight into the thermal
dynamics. To address the behaviour of systems of mutually interacting
particles, we demonstrate that results for the relevant shift and
hypercurrent correlation functions are accessible in molecular
dynamics.

The access to the gauge correlation functions in molecular simulations
is provided by implementing differentiation with respect to the
initial state.  Automatic differentiation \cite{baydin2018autodiff} is
a natural choice for realizing this operation, and it has gained much
recent popularity, as it allows one to access with great ease
derivatives that are otherwise out of practical reach of symbolic
differentiation, whether via pencil and paper or algorithmically
assisted. The method does not suffer from the drawbacks of numerical
finite-difference schemes and it has been used to great effect in
molecular dynamics simulations in recent work addressing design and
optimal control tasks \cite{krueger2024prl, king2024pnas,
  engel2023prx, schoenholz2020}, as well as in the implementation of
functional calculus based on analytical \cite{stierle2024autodiff,
  sammueller2023whyNeural} and machine-learned neural density
functionals \cite{sammueller2023neural, sammueller2023whyNeural,
  sammueller2024hyperDFT, sammueller2024whyhyperDFT,
  sammueller2024pairmatching, sammueller2024attraction,
  buchannan2025attraction, kampa2024meta, bui2024neuralrpm,
  dijkman2024ml, glitsch2025disks, yang2024}.  Here we exploit that
the dynamical gauge invariance is inherently linked to a specific
initial state derivative of the nonequilibrium dynamics. We also
demonstrate the alternative accessibility of all gauge correlation
functions via finite-difference differentiation, which works
universally with no need for specialized compute environment.

The paper is organized as follows. In
Sec.~\ref{SECstatisticalMechanics} we describe the particle-based
classical statistical mechanics, including the setup for the
Hamiltonian in Sec.~\ref{SECnonequilibriumDynamics} and the
Liouvillian formulation of the dynamics in
Sec.~\ref{SECtimeEvolution}. We lay out the microscopically resolved
one-body equation of motion in Sec.~\ref{SEConeBodyLevel} and describe
the standard realization of the dynamics via many-body trajectories in
Sec.~\ref{SECtrajectories}.

The dynamical gauge invariance theory is developed in
Sec.~\ref{SECdynamicalGaugeInvariance}, starting with an elementary
derivation of an exact dynamical shift current sum rule in
Sec.~\ref{SECdynamicalHyperforceCorrelations}. We then address the
deeper mechanism underlying this derivation by formalizing the phase
space shifting on the basis of Poisson brackets in
Sec.~\ref{SECshiftingViaPoissonBrackets}. The Poisson bracket
formulation is then used to generalize to dynamical phase space
shifting and thus to formulate dynamical gauge invariance in
Sec.~\ref{SECdynamicalPhaseSpaceShifting}, which allows one to derive
exact hypercurrent sum rules that apply to general observables under
nonequilibrium dynamics, as described in
Sec.~\ref{SEChyperobersableSumRules}.  The connection to a specific
initial-state time derivative is presented in
Sec.~\ref{SECinitialStateTimeDifferentiation}.  The relevance for
trajectory-based simulations is described in
Sec.~\ref{SECmolecularDynamicsConcepts}. We illustrate the general
theory by making specific choices for relevant hyperobservables in
Sec.~\ref{SECconcreteSumRules}.

We present applications of the general gauge theory in
Sec.~\ref{SECapplications}.  We first demonstrate that in equilibrium
the formalism recovers correctly the static hyperforce theory
\cite{robitschko2024any, mueller2024gauge, mueller2024whygauge}, while
providing additional nontrivial temporal insight, as presented in
Sec.~\ref{SEClinkToEquilibrium}.  As an initial concrete toy model
that permits analytical solution, we consider harmonically confined
noninteracting particles in Sec.~\ref{SECharmonicOscillators}. To
address mutually interacting systems, we turn to simulations and use
molecular dynamics together with initial state differentiation
implemented via automatic or finite-difference derivatives to access
the shift current and hypercurrent correlation functions in
Sec.~\ref{SECmolecularDynamicsResults}. The framework allows one to
address the construction of nonequilibrium reduced-variance estimators
and to shed new light on the classical barrier crossing problem.  We
give conclusions in Sec.~\ref{SECconclusions}.

\section{Statistical Mechanics}
\label{SECstatisticalMechanics}

\subsection{Microscopic many-body model}
\label{SECnonequilibriumDynamics}

We consider classical systems of $N$ particles with position
coordinates $\rv_1,\ldots,\rv_N\equiv \rv^N$ and momentum variables
$\pv_1,\ldots, \pv_N\equiv\pv^N$, where $\rv^N$ and $\pv^N$ are
shorthand notations. The Hamiltonian is taken to possess the standard
form
\begin{align}
  H = \sum_i \frac{\pv_i^2}{2m} + u(\rv^N) + \sum_i V_\rmext(\rv_i),
  \label{EQHamiltonian}
\end{align}
where the summation index $i=1,\ldots,N$ runs over all $N$ particles,
$m$ denotes the particle mass, $u(\rv^N)$ is the interparticle
interaction potential, and $V_\rmext(\rv)$ is an external potential,
here expressed as a function of the generic position variable
$\rv$. The initial state at times $t<0$ is characterized by a
stationary Hamiltonian~$H_0$, where the mass $m_0$, interparticle
potential $u_0(\rv^N)$, and external potential $V_{\rmext,0}(\rv)$ are
all explicitly time-independent, as is indicated by the subscript 0.
Per construction the initial system possesses a well-defined thermal
equilibrium, as described below.  For times $t \geq 0$, general
explicit time-dependence in \eqr{EQHamiltonian} can occur. This
includes temporal variation of the mass $m$, of the interparticle
interaction potential $u(\rv^N)$, and of the external potential
$V_\rmext(\rv)$; here and throughout we suppress mere {\it parametric}
dependence on time $t$ in the notation and reserve all time arguments
to denote {\it dynamical} dependence, as it is induced by the particle
motion.

The switching at $t=0$ need not be smooth and hence a discontinuity
may occur such that in general we allow for: $\lim_{t\to 0^+}H \neq
H_0$.  That the initial state Hamiltonian $H_0$ and the nonequilibrium
time evolution generator $H$ can in general differ from each other
allows for flexible modelling of a broad range of situations. Thereby
the initial many-body phase space probability distribution
$f_0(\rv^N,\pv^N)$ is taken to be in thermal equilibrium, as
characterized by the canonical ensemble for the given form of
$H_0$. Hence as a function of the phase space point $\rv^N, \pv^N$ we
have:
\begin{align}
  f_0(\rv^N, \pv^N) &= \frac{\e^{-\beta H_0}}{Z_0},
  \label{EQf0}
\end{align}
where $\beta=1/(k_BT)$, with Boltzmann constant $k_B$, absolute
temperature $T$ of the initial state, and the (scaled) canonical
partition sum $Z_0=\Tr \e^{-\beta H_0}$.  The normalization is such
that $\Tr f_0=1$, where the canonical `trace' is the phase space
integral over all degrees of freedom: $\Tr \cdot = \int d\rv_1\ldots
d\rv_N d\pv_1\ldots d\pv_N \cdot$. The partition sum in standard form
is then $Z_0/(h^{dN}N!)$ with Planck constant $h$ and dimensionality
$d$.

As a special case, the present dynamical setup includes the switching
at time $t=0$ from an initial Hamiltonian $H_0$ to a different
Hamiltonian $H\neq H_0$, where $H$ however is also taken to be
stationary. Then the system characterized by $H$ is driven out of {\it
  its} associated equilibrium already at the initial time, which forms
a common situation, e.g.\ when applying linear response
theory~\cite{zwanzig2001}.

\subsection{Phase space time evolution and dynamical averages}
\label{SECtimeEvolution}

The time evolution of the dynamical probability distribution function
$f(\rv^N,\pv^N,t)$ is governed by the Liouville equation $\partial
f(t)/\partial t = -Lf(t)$, where we have left away the dependence on
the phase space variables $\rv^N, \pv^N$ in the notation. The
Liouvillian $L$ is the phase space differential operator given via
Poisson brackets by:
\begin{align}
  L=\{\,\cdot\,, H\}.
  \label{EQLiouvillian}
\end{align}
The Poisson bracket $\{\,\cdot\,,\,\cdot\,\}$ of two general phase
space functions $g$ and $h$ is thereby defined \cite{goldstein2002}
as:
\begin{align}
\{g, h\} = \sum_i \Big(
  \frac{\partial g}{\partial\rv_i}\cdot
  \frac{\partial h}{\partial \pv_i} - 
  \frac{\partial g}{\partial \pv_i} \cdot
  \frac{\partial h}{\partial \rv_i} \Big).
\end{align}

In our chosen setup, the time evolution for $t< 0$ is governed by the
stationary Liouvillian $L_0=\{\,\cdot\,, H_0\}$, where we recall that
$H_0$ carries no explicit time dependence {and that in general
  $H_0\neq H(0^+)$}.  At later times $t\geq 0$ any potentially
occurring parametric dependence of $H$ on time~$t$ is passed on to $L$
in \eqr{EQLiouvillian}; as noted above we suppress the parametric
dependence on time throughout in the notation, such that time
arguments denote exclusively the dynamical dependence.

Due to the specific additive structure of the Hamiltonian
\eqref{EQHamiltonian}, the Liouvillian $L$ consists correspondingly of
a sum of ideal (kinetic), interparticle, and external contributions
according to
\begin{align}
  L=L_\rmid + L_\rmint + L_\rmext.
  \label{EQLsplitting}
\end{align}
Inserting the general form of the Hamiltonian \eqref{EQHamiltonian}
into the generic Poisson bracket form of the Liouvillian
\eqref{EQLiouvillian} gives explicit expressions for the three
contributions in the Liouvillian splitting
\eqref{EQLsplitting}. Thereby the ideal part is $L_\rmid = \sum_i
(\pv_i/m)\cdot\nabla_i$ as is generated by the kinetic energy in
\eqr{EQHamiltonian}, the interparticle part 
is $L_\rmint = -\sum_i [\nabla_i u(\rv^N)]\cdot\nabla_{\pv_i}$, and
the external contribution is $ L_\rmext = -\sum_i [\nabla_i
  V_\rmext(\rv_i)]\cdot \nabla_{\pv_i}$, and the latter two forms
arise, respectively, from the interparticle potential $u(\rv^N)$ and
external potential energy $\sum_i V_\rmext(\rv_i)$ in
\eqr{EQHamiltonian}. Forces are generated from position gradients,
where $\nabla_i=\partial/\partial \rv_i$ denotes the derivative with
respect to position~$\rv_i$ and
$\nabla_{\pv_i}=\partial/\partial\pv_i$ is the partial derivative with
respect to the momentum of particle~$i$.

The Liouvillian splitting \eqref{EQLsplitting} into its constituent
parts follows a standard scheme. A prominent example is the derivation
of the celebrated velocity Verlet algorithm \cite{verlet1967,
  frenkel2023book}, starting from the Trotter expansion of the
propagator \cite{hansen2013}, which we briefly relate to.  In the
notation of Ref.~\cite{hansen2013}, the Liouvillian is decomposed as
$L = {\rm i}{\cal L}_\rv + {\rm i}{\cal L}_\pv$, with the position
contribution ${\rm i}{\cal L}_\rv = L_\rmid$ and the momentum part
${\rm i}{\cal L}_\pv = L_\rmint + L_\rmext$, where by convention the
imaginary unit~${\rm i}$ is included \cite{schilling2022}. This
further splitting of ${\rm i}{\cal L}_\pv$ into interparticle
($L_\rmint$) and external contributions ($L_\rmext$) follows naturally
from the additive form of the Hamiltonian~\eqref{EQHamiltonian}.

Formally, the many-body probability distribution $f(t)$ that solves
the Liouville equation can be expressed as
\begin{align}
  f(t)=\calU(t,0)f_0, 
 \label{EQfdynamicalFromU}
 \end{align}
where $\calU(t,0)$ denotes the propagator that performs the time
evolution from time 0 to time $t$. The propagator can be expressed as
$\calU(t,0)=\e_+^{\int_0^t dt' L'}$, where $\e_+$ denotes the
positively time-ordered exponential \cite{sakurai1973book} and $L'$
denotes the Liouvillian \eqref{EQLiouvillian}, which is {\it
  parametrically} evaluated at time~$t'$. Per construction
$\calU(0,0)=1$ such that $f(0)=\calU(0,0) f_0$. Working with the
propagator $\calU(t,0)$ is a powerful formal method that we rely on in
the following. The equivalent and arguably more intuitive
trajectory-based picture of the Hamiltonian dynamics is laid out below
in Sec.~\ref{SECtrajectories}.

Any general phase space function $\hat A(\rv^N, \pv^N)$ that acts as
an observable of interest acquires dynamical time dependence, as
generated from the motion in the system, via
\begin{align}
  \hat A(t) = \calU^\dagger(t,0) \hat A,
  \label{EQhatAdynamicalFromU}
\end{align}
where the dagger indicates the adjoint; here the adjoint of an
operator ${\cal O}$ is defined in the standard way as: $\Tr g {\cal O}
h = \Tr h {\cal O}^\dagger g$, where we recall $g$ and $h$ being two
phase space functions and $\Tr$ indicating the full phase space
integral.  The propagator $\calU(t,0)$ and its adjoint
$\calU^\dagger(t,0)$ are inverse of each other, such that
$\calU(t,0)\calU^\dagger(t,0)= 1$ and
$\calU^\dagger(t,0)\calU(t,0)=1$, as is characteristic of unitary time
evolution.

We have left away the phase space arguments for brevity of
notation. More explicitly, the variable $\hat A(\rv^N, \pv^N)$ on the
right hand side of \eqr{EQhatAdynamicalFromU} represents both the
observable itself as well as its value at the initial microstate
$\rv^N, \pv^N$ \cite{zwanzig2001}.  Using a quantum analogy, $\hat A$
is a Schr\"odinger observable. Then the corresponding dynamical
Heisenberg observable $\hat A(\rv^N,\pv^N,t)$ is given via
\eqr{EQhatAdynamicalFromU}. In the present classical context $\hat
A(\rv^N,\pv^N,t)$ constitutes the value that the observable attains at
time $t$, given the specific trajectory that started at $\rv^N, \pv^N$
at the initial time \cite{zwanzig2001}, as is laid out further in
Sec.~\ref{SECtrajectories}. Hence $\hat A(t)=\hat A(\rv^N,\pv^N,t)$
remains dependent on the initial state, which will be important in the
development of the dynamical gauge theory in
Sec.~\ref{SECdynamicalGaugeInvariance}.

The Liouville equation for general Heisenberg observables is $\partial
\hat A(t)/\partial t = L(t) \hat A(t)$, where the time-evolved
Liouvillian $L(t)$ is obtained from $L$, as defined via
\eqr{EQLiouvillian}, by:
\begin{align}
  L(t) = \calU^\dagger(t,0) L \calU(t,0).
  \label{EQLoft}
\end{align}
An alternative form to \eqr{EQLoft} is obtained by using Poisson
brackets: $L(t) = \{\,\cdot\,, H(t)\}$, where the temporally evolved
form of the Hamiltonian is obtained according to the standard form of
a Heisenberg observable as $H(t)= \calU^\dagger(t,0)H$, with all
parametric time dependences being evaluated at time $t$ and suppressed
in the notation.  Time-resolved averages are then obtained as $A(t) =
\langle \hat A(t) \rangle = \Tr f_0 \hat A(t)$, which can equivalently
be written as $A(t)=\Tr f(t) \hat A$.

The present formalism for the dynamics on the full many-body level
requires one to choose a coarse-graining strategy to proceed in
formulating a statistical mechanical description.

\subsection{One-body level of correlation functions}
\label{SEConeBodyLevel}

We work with microscopic resolution, using the classical operators
(phase space functions) for the one-body density, $\hat\rho(\rv)
=\sum_i\delta(\rv-\rv_i)$, and for the one-body current,
\begin{align}
  \hat\Jv(\rv) &= \sum_i \delta(\rv-\rv_i)\frac{\pv_i}{m},
  \label{EQcurrentOperator}
\end{align}
where $\delta(\cdot)$ denotes the Dirac distribution in $d$
dimensions. We first lay out the standard derivation of the dynamical
one-body force density balance, which constitutes Newtons' second law
in the present context \cite{schmidt2022rmp}. The Heisenberg
observable for the current (momentum density) is obtained as $m\hat
\Jv(\rv,t) = \calU^\dagger(t,0)m\hat\Jv(\rv)$ and it satisfies the
Liouville equation of motion $\partial m\hat\Jv(\rv,t)/\partial t =
L(t) m\hat\Jv(\rv,t)$. Explicit calculation of the latter right hand
side yields the Heisenberg force density observable:
\begin{align}
  \hat \Fv(\rv,t) &= L(t) m \hat\Jv(\rv,t).
  \label{EQforceDensityOperator}
\end{align}
where $\hat\Fv(\rv,t) = \calU^\dagger(t,0)\hat\Fv(\rv)$. Using the
splitting \eqref{EQLsplitting} of the Liouvillian, the force density
$\hat\Fv(\rv)$ can be decomposed correspondingly into ideal (or
kinetic), interparticle, and external parts: $\hat \Fv(\rv)=
\nabla\cdot\hat\taub(\rv) + \hat\Fv_\rmint(\rv) - \hat\rho(\rv) \nabla
V_\rmext(\rv)$, with the kinetic stress operator $\hat\taub(\rv) =
-\sum_i \delta(\rv-\rv_i) \pv_i\pv_i/m$, the interparticle force
density operator $\hat \Fv_\rmint(\rv) = -\sum_i
\delta(\rv-\rv_i)\nabla_i u(\rv^N)$, and the external force field
$-\nabla V_\rmext(\rv)$ and we recall that $\hat\rho(\rv)$ denotes the
microsopically-resolved density operator defined inline above
\eqr{EQcurrentOperator}.  As before, any parametric dependence of $m,
u(\rv^N)$, and $V_\rmext(\rv)$ on time~$t$ remains suppressed here and
throughout in the notation.

We address the dynamics by considering $\partial
m\hat\Jv(\rv,t)/\partial t =$ $ L(t) m\hat \Jv(\rv,t)
=\calU^\dagger(t,0) L m\hat\Jv(\rv) =\calU^\dagger(t,0)\hat\Fv(\rv)$,
where we have first written out $L(t)$ via \eqr{EQLoft}, then used the
identity $\calU(t,0)\calU^\dagger(t,0)=1$ and have identified $\hat
\Fv(\rv) = L m\hat\Jv(\rv)$.  Averaging over the initial distribution
$f_0$ and using the decomposition \eqref{EQLsplitting} then yields the
one-body equation of motion \cite{schmidt2022rmp}:
\begin{align}
  \Fv_\rmid(\rv,t) + \Fv_\rmint(\rv,t) + \Fv_\rmext(\rv,t) 
  &= \frac{\partial m\Jv(\rv,t)}{\partial t},
  \label{EQofMotionJ}
\end{align} 
where the ideal (kinetic) force density is $\Fv_\rmid(\rv,t) = \nabla
\cdot \langle \hat \taub(\rv,t) \rangle$, the interparticle force
density is $\Fv_\rmint(\rv,t) = \langle \hat
\Fv_\rmint(\rv,t)\rangle$, and the external force density is
$\Fv_\rmext(\rv,t) = -\rho(\rv,t)\nabla V_\rmext(\rv,t)$. Here the
dynamical density profile is $\rho(\rv,t)=\langle \hat\rho(\rv,t)
\rangle$ and the mean one-body current on the right hand side of
\eqr{EQofMotionJ} is $\Jv(\rv,t) = \langle \hat \Jv(\rv,t) \rangle$
with the Heisenberg current observable given below
\eqr{EQcurrentOperator}. We recall that averages are built in the
standard way as $\langle \,\cdot\, \rangle=\Tr f_0 \,\cdot\,$, see
Sec.~\ref{SECtimeEvolution}.

Turning to the initial system, which is in thermal equilibrium
according to $f_0$ given by \eqr{EQf0}, the general dynamical force
density balance \eqref{EQofMotionJ} reduces to the static force
density relationship $\Fv_0(\rv)=0$, which can be decomposed as
\begin{align}
  \Fv_{\rmid,0}(\rv) + \Fv_{\rmint,0}(\rv)
  + \Fv_{\rmext,0}(\rv) &= 0,
  \label{EQforceDensityBalanceStatic}
\end{align}
where explicitly the three terms are: $\Fv_{\rmid,0}(\rv)=\nabla \cdot
\Tr f_0 \hat\taub(\rv)$, $\Fv_{\rmint,0}(\rv)=-\Tr f_0 \sum_i\delta
(\rv-\rv_i)\nabla_i u_0(\rv^N)$, and $\Fv_{\rmext,0}(\rv)=
-\rho_0(\rv)\nabla V_{\rmext,0}(\rv)$, with the initial state density
profile $\rho_0(\rv)= \Tr f_0 \hat\rho(\rv)$. The static force density
balance \eqref{EQforceDensityBalanceStatic} is often refered to as
Yvon-Born-Green equation \cite{yvon1935, born1946}, in particular for
systems that interact by pairwise interparticle force only, such that
$\Fv_{\rmint,0}(\rv)$ can be written as an integral over the two-body
density distribution \cite{hansen2013, evans1979}.

\subsection{Trajectories and initial-state differentiation}
\label{SECtrajectories}

The Liouvillian dynamics described in Sec.~\ref{SECtimeEvolution} can
be complemented usefully by a formulation that is more explicitly
based on trajectories. This setup can both guide intuition and provide
a practical route to constructing analytical solutions and performing
simulation work, as will prove useful below.  When working with
individual trajectories, the two relevant tasks are to realize initial
states according to the distribution function $f_0(\rv^N, \pv^N)$ and
further to obtain a representation of the trajectories $\hat\rv_i(t),
\hat\pv_i(t)$ for all particles $i=1,\ldots, N$. While for very simple
systems analytical work is possible (an example is presented in
Sec.~\ref{SECharmonicOscillators}), in general and for realistic
systems of interest one needs to rely on simulations.

Generating the initial distribution of microstates~$f_0(\rv^N, \pv^N)$
according to the canonical ensemble form \eqref{EQf0} is a standard
task, which can be efficiently performed via Monte Carlo simulations
for the given choice of the initial Hamiltonian $H_0$ and prescribed
value of the inverse temperature $\beta$. We recall that the choice of
initial Hamiltonian is independent of the subsequent time evolution
based on $H$, where in general $H \neq H_0$.  Then, to address the
dynamics the verlocity Verlet algorithm~\cite{hansen2013,
  frenkel2023book} is an apt choice to integrate Hamilton's equations
of motion and hence to obtain trajectories $\hat\rv_i(\rv^N,\pv^N,t)$
and $\hat\pv_i(\rv^N, \pv^N, t)$ for all particles $i=1,\ldots, N$,
starting from the initial state $\rv^N, \pv^N$.

General Heisenberg observables $\hat A(t)$ can then be represented by
$\hat A(\rv^N,\pv^N,t) = \hat A (\hat\rv^N(t), \hat\pv^N(t))$, where
we recall that the phase space point $\rv^N, \pv^N$ on the left hand
side is the initial state.  On the right hand side the Schr\"odinger
observable $\hat A$ is evaluated at the phase space point
$\hat\rv^N(t), \hat\pv^N(t)$, which is the configuration that the
system has reached at time~$t$.

\begin{figure}[!t]
  \vspace{1mm}
  \includegraphics[page=1,width=.99\columnwidth]{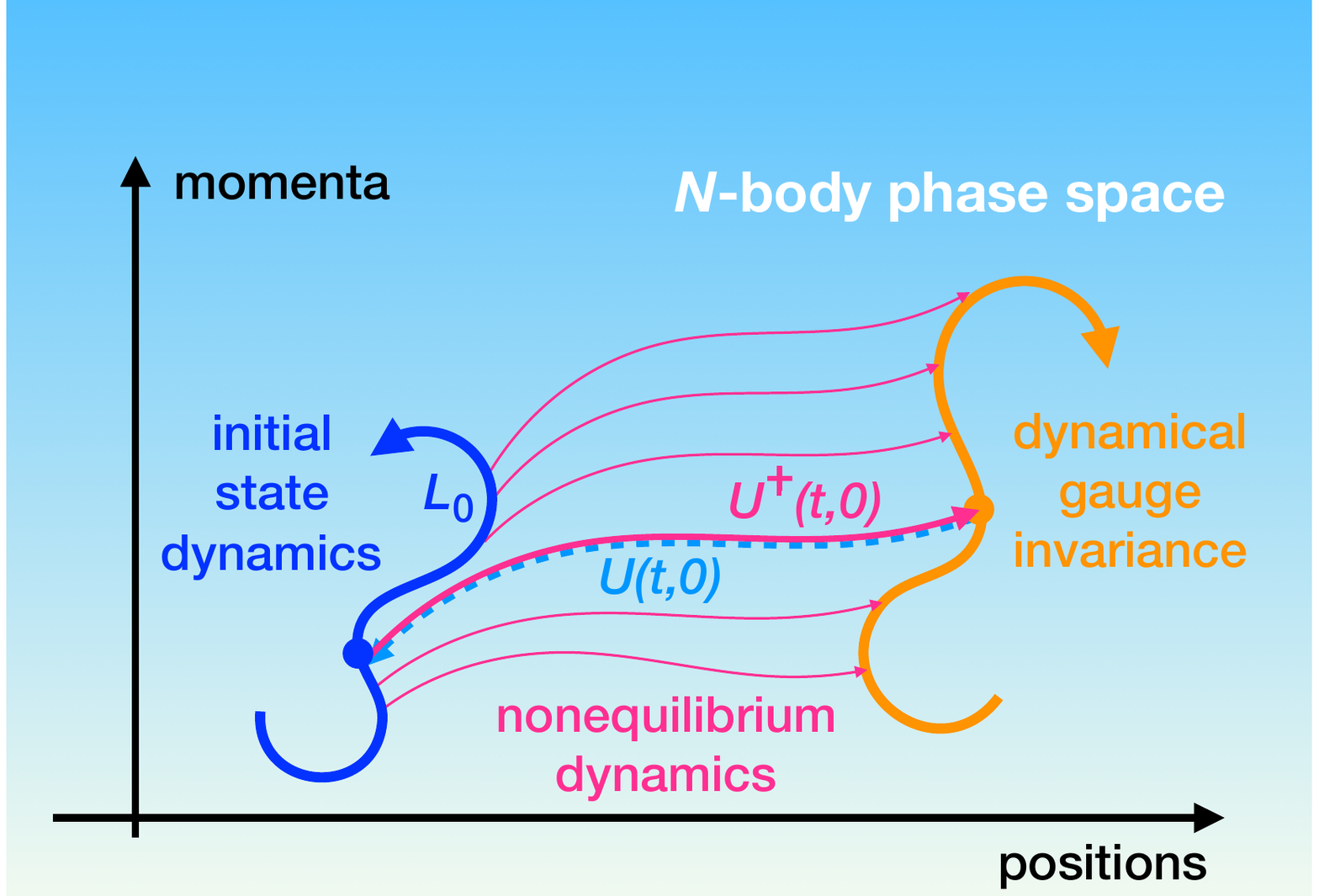}
  \caption{Illustration of the different types of time evolution as
    represented by $N$-body trajectories on phase space. The initial
    state dynamics are governed by the initial state Liouvillian
    $L_0$, as represented by one specific trajectory (dark blue curve
    on the left). Each point on this trajectory serves as an initial
    state for the nonequilibrium time evolution (magenta arrows in the
    middle). These dynamics are transported by the adjoint propagator
    $\calU^\dagger(t,0)$ forward in time and $\calU(t,0)$ performs the
    inverse operation (light blue dashed arrow to the left). The
    initial state trajectory is mapped via $\calU^\dagger(t,0)$ onto a
    specific trajectory (orange curve on the right) along which
    dynamical gauge invariance applies.}
\label{FIG2}
\end{figure}

As an important computational point, we demonstrate below that this
interpretation gives immediate and practical access to derivatives
with respect to the initial state. This will be key for evaluating the
arising gauge correlation functions, as laid out in
Sec.~\ref{SECdynamicalGaugeInvariance}, in simulation work. In
particular, the task within the gauge correlation framework is to
implement a specific initial-state time derivative according to the
action of the initial-state Liouvillian $L_0$, see below for further
details. In practice, the differentiation can be performed via a
finite-difference scheme by altering the state according to the
initial state dynamics and rerunning the simulation. As a numerically
robust alternative, the powerful methodology of automatic
differentiation can be leveraged. This technique gives access to the
$(2Nd) \times (2Nd)$ Jacobian matrix $\partial (\hat{\rv}^N(t),
\hat{\pv}^N(t) )/\partial( \rv^N, \pv^N)$, from which the
initial-state time derivative follows naturally as further laid out in
Sec.~\ref{SECinitialStateTimeDifferentiation}.  An illustration of the
underlying concepts is shown in Fig.~\ref{FIG2}.

\section{Dynamical gauge invariance}
\label{SECdynamicalGaugeInvariance}

\subsection{Exact shift current sum rule as a prototype}
\label{SECdynamicalHyperforceCorrelations}

We complement the standard dynamical one-body picture described in
Sec.~\ref{SEConeBodyLevel} by rather letting the {\it initial time}
evolution act on the current operator at time $t$.  Hence we apply the
thermally scaled initial state Liouvillian $\beta L_0$ to the
nonequilibrium phase space current:
\begin{align}
  \hat\Cv(\rv,t) &= \beta  L_0 m\hat \Jv(\rv,t).
  \label{EQdefinitionOfBare}
\end{align}
We thereby recall the Schr\"odinger definition
\eqref{EQcurrentOperator} of the current observable $\hat\Jv(\rv)$ and
that the dependence on the phase space point, suppressed in the above
notation of the Heisenberg current $\hat\Jv(\rv,t)$, describes the
initial microstate $\rv^N, \pv^N$ on which $\beta L_0$ acts in
\eqr{EQdefinitionOfBare}.

As we demonstrate, the present strategy is both conceptually and in
practice distinct from the conventional time evolution generated by
$L(t)$, thus leading to new insight that emerges from introducing the
classical `shift current' observable $\hat \Cv(\rv,t)$ via
\eqr{EQdefinitionOfBare}.  As an illustration, we spell this out more
explicity, using $\hat\Jv(\rv,t)=\calU^\dagger(t,0)\hat\Jv(\rv)$ as
the temporally evolved current~\eqref{EQcurrentOperator}, which upon
insertion into \eqr{EQdefinitionOfBare} yields $\hat \Cv(\rv,t)= \beta
L_0 \calU^\dagger(t,0)\sum_i \delta(\rv-\rv_i)\pv_i$. The latter form
makes explicit: i) that the mass $m$ cancels out from the definition
of $\hat \Cv(\rv,t)$; ii) that the initial state Liouvillian $L_0$
acts {\it after} the adjoint propagator $\calU^\dagger(t,0)$ does --
we return to the significance of this temporal structure when
discussing the trajectory level in
Sec.~\ref{SECmolecularDynamicsConcepts}; and iii) that at the initial
time $t=0$ the shift current reduces to the scaled force density,
$\hat\Cv(\rv,0) = \beta \hat\Fv_0(\rv)$, as follows from
\eqr{EQforceDensityOperator} and where again the subscript 0 indicates
the application to the initial state.

We turn to the statistical mechanical consequences that are implied by
the shift current observable \eqref{EQdefinitionOfBare}. The mean
one-body shift current $\Cv(\rv,t)$ follows via averaging
\eqr{EQdefinitionOfBare} according to $\Cv(\rv,t) =\langle \hat
\Cv(\rv,t) \rangle$ $=\Tr f_0 \beta L_0 m\hat\Jv(\rv,t)$. Spelling out
the Poisson bracket form of the initial state Liouvillian
$L_0=\{\,\cdot\,,H_0\}$ gives $\Cv(\rv,t) = \Tr f_0 \{m\hat\Jv(\rv,t),
\beta H_0\}$.  Applying the chain rule to the exponential form of the
initial state equilibrium distribution function $f_0(\rv^N, \pv^N)$
leads to $\Cv(\rv,t) = -\Tr\{m\hat\Jv(\rv,t), f_0\}=0$. That the
result vanishes follows from the general property of the phase space
integral of any Poisson bracket vanishing, $\Tr \{g,h\}=0$, as is
readily seen via integration by parts on phase space (for $g$ and $h$
being well behaved).

In summary, we have shown that the average of the shift current
\eqref{EQdefinitionOfBare} vanishes,
\begin{align}
  \Cv(\rv,t)=0,
  \label{EQshiftingCurrentVanishes}
\end{align}
which is an exact nonequilibrium sum rule.  We express this identity
by splitting the left hand side into three contributions:
\begin{align}
  \Cv_\rmid(\rv,t) + \Cv_\rmint(\rv,t) + \Cv_\rmext(\rv,t)   &= 0,
  \label{EQbareHyperForceBalance}
\end{align}
where the decomposition into the three terms follows from the
corresponding splitting \eqref{EQLsplitting} of the initial state
Liouvillian $L_0$; we recall the presence of $L_0$ in the definition
\eqref{EQdefinitionOfBare} of the shift current observable. As the
splitting is crucial in what follows we spell out the details, first
recalling that the initial Hamiltonian $H_0$ consists of kinetic,
interparticle and external parts according \eqr{EQHamiltonian}.

The splitting \eqref{EQbareHyperForceBalance} arises then from
decomposing the initial state Liouvillian $L_0$ according to
\eqref{EQLsplitting}, such that $L_0 = L_{\rmid,0} + L_{\rmint,0} +
L_{\rmext,0}$ in \eqr{EQdefinitionOfBare}, thereby using the initial
particle mass $m_0$, interparticle potential $u_0(\rv^N)$ and external
potential $V_{\rmext,0}(\rv)$.  The corresponding ideal,
interparticle, and external shift currents are
$\Cv_\alpha(\rv,t)=\langle \hat \Cv_\alpha(\rv,t) \rangle = \langle
\beta L_{\alpha,0} m \hat \Jv(\rv,t)\rangle$, where $\alpha =$
`$\rmid$', `$\rmint$', and `$\rmext$', respectively.  At the initial
time, $t=0$, the shift current balance \eqref{EQbareHyperForceBalance}
reduces to the equilibrium force density balance
\eqref{EQforceDensityBalanceStatic}; we give an extended account of
this limit below in Sec.~\ref{SEClinkToEquilibrium}.

It is revealing to compare the structure of the exact dynamical sum
rule \eqref{EQbareHyperForceBalance} with that of the one-body
equation of motion~\eqref{EQofMotionJ}.  Both equations are vectorial,
and their left hand sides share an analogous splitting into ideal,
interparticle, and external contributions. However, the right hand
side of the equation of motion \eqref{EQofMotionJ} is the time
derivative of the current, which in general is very different from the
universal zero on the right hand side of the nonequilibrium sum rule
\eqref{EQbareHyperForceBalance}.  Before demonstrating the validity of
the shift current balance~\eqref{EQbareHyperForceBalance} below, we
first aim to uncover the fundamental mechanism that lies behind the
above derivation.

\subsection{Static phase space shifting via Poisson brackets}
\label{SECshiftingViaPoissonBrackets}
The derivation of the shift current balance
\eqref{EQshiftingCurrentVanishes} and thus of its split form
\eqref{EQbareHyperForceBalance} in
Sec.~\ref{SECdynamicalHyperforceCorrelations} relies on both the
one-body localization and on the properties of the Poisson brackets;
the latter occur in the initial state Liouvillian $L_0$; recall
\eqr{EQLiouvillian}. It is hence natural to use the Poisson brackets
as a foundation for defining the following spatially localized
differential operators:
\begin{align}
  \bsig(\rv) &= \{\,\cdot\,, m \hat\Jv(\rv)\},
  \label{EQbsigAsPoissonBracket}
\end{align}
where we recall that $\hat\Jv(\rv)$ is the microscopically given
current phase space function \eqref{EQcurrentOperator} and thus
$m\hat\Jv(\rv)=\sum_i\delta(\rv-\rv_i)\pv_i$. Applying the
differential operator \eqref{EQbsigAsPoissonBracket} to the (negative)
Hamiltonian $H$ yields the classical force density operator
$\hat\Fv(\rv)$ according to
\begin{align}
  -\bsig(\rv) H = \{m\hat \Jv(\rv), H \} = L m\hat\Jv(\rv)
  = \hat \Fv(\rv),
  \label{EQforceDensityOperatorFrombsig}
\end{align}
where we have first written out the Poisson bracket
\eqref{EQbsigAsPoissonBracket} and then identified both the
Liouvillian \eqref{EQLiouvillian} and the classical force density
operator $\hat \Fv(\rv)$ according to \eqr{EQforceDensityOperator}.
As an aside we emphasize that no dynamical dependence is implied yet
in \eqr{EQforceDensityOperatorFrombsig}. The present Schr\"odinger
force density $\hat \Fv(\rv)$ is related to the corresponding
dynamical Heisenberg observable via $\hat \Fv(\rv,t) =
\calU^\dagger(t,0) \hat\Fv(\rv)$, see \eqr{EQforceDensityOperator}.

The properties of the Poisson bracket render the differential
operators $\bsig(\rv)$ anti-self-adjoint on phase space, $\bsig(\rv)=
-\bsig^\dagger(\rv)$, as follows via integration by parts on phase
space \cite{mueller2024gauge}. The explicit form of $\bsig(\rv)$ is
obtained straightforwardly by inserting the definition
\eqref{EQcurrentOperator} of the classical current operator
$\hat\Jv(\rv)$ into the Poisson
bracket~\eqref{EQbsigAsPoissonBracket}, which yields
$\bsig(\rv)=\{\,\cdot\,,\sum_i \pv_i\delta(\rv-\rv_i)\}$.  Simplifying
gives the explicit form:
\begin{align}
\bsig(\rv) = \sum_i
[\delta(\rv-\rv_i)\nabla_i + \pv_i
  \nabla\delta(\rv-\rv_i)\cdot\nabla_{\pv_i}],
\label{EQbsigExplicit}
\end{align}
which reveals $\bsig(\rv)$ to be {\it identical} to the localized
differential operators that represent the static `shifting' gauge
invariance of {\it equilibrium} statistical
mechanics~\cite{mueller2024gauge, mueller2024whygauge}.

The immediate implications are: i)~the validity of non-trivial
commutator structure of $\bsig(\rv)$ and $\bsig(\rv')$ at positions
$\rv,\rv'$ \cite{mueller2024gauge, mueller2024whygauge} and ii)~the
geometric interpretation via phase space shifting according to
particle position displacement and corresponding momentum
transform. These general properties continue to hold for the Poisson
bracket form \eqref{EQbsigAsPoissonBracket} as they follow directly
from the structure of phase space \cite{mueller2024gauge,
  mueller2024whygauge}, irrespective of whether a static
\cite{mueller2024gauge, mueller2024whygauge} or the present dynamical
statistical mechanical setup is considered.  We recall the pertinence
of the infinitesimal shifting operators for generating equilibrium sum
rules involving forces and hyperforce correlation functions
\cite{robitschko2024any, mueller2024gauge, mueller2024whygauge}
according to the mechanism in~\eqr{EQforceDensityOperatorFrombsig}.

\subsection{Dynamical phase space shifting operators}
\label{SECdynamicalPhaseSpaceShifting}

We generalize the static shifting operators
\eqref{EQbsigAsPoissonBracket} to dynamical counterparts
$\bsig(\rv,t)$ according to the standard mechanism to transform
differential operators:
\begin{align}
  \bsig(\rv,t) &= \calU^\dagger(t,0)\bsig(\rv)\calU(t,0),
  \label{EQsigmaDynamicalAsPropagatorSandwich}
\end{align}
where we recall that $\calU(t,0)$ is the (phase space) propagator from
the initial time 0 to time $t$ and $\calU^\dagger(t,0)$ denotes its
(phase space) adjoint.  Mirroring the properties of the static version
$\bsig(\rv)$, the dynamical shifting operators are anti-self-adjoint
on phase space:
\begin{align}
  \bsig(\rv,t) &= -\bsig^\dagger(\rv,t).
  \label{EQbsigDynamicalAntiSelfAdjoint}
\end{align}

The proof of \eqr{EQbsigDynamicalAntiSelfAdjoint} follows
straightforwardly via explicit calculation:
$[\calU^\dagger(t,0)\bsig(\rv)\calU(t,0)]^\dagger=
\calU^\dagger(t,0)\bsig^\dagger(\rv)\calU^{\dagger\dagger}(t,0)=-\bsig(\rv,t)$,
where we have first built the overall adjoint by reversing the factors
and then have used both $\bsig^\dagger(\rv)=-\bsig(\rv)$ and
$\calU^{\dagger\dagger}(t,0)=\calU(t,0)$, which allows one to identify
the final result via \eqr{EQsigmaDynamicalAsPropagatorSandwich}.  As
an aside, multiplying \eqr{EQsigmaDynamicalAsPropagatorSandwich} by
$\calU^\dagger(t,0)$ from the right yields the identity
$\bsig(\rv,t)\calU^\dagger(t,0) = \calU^\dagger(t,0)\bsig(\rv)$ and
corresponding multiplication by $\calU(t,0)$ from the left yields
$\calU(t,0) \bsig(\rv,t) = \bsig(\rv) \calU(t,0)$.

As an alternative to the propagator `sandwich'
form~\eqref{EQsigmaDynamicalAsPropagatorSandwich}, the dynamical
shifting operators can be expressed equivalently as
\begin{align}
  \bsig(\rv,t) &= \{\,\cdot\,, m\hat\Jv(\rv,t)\}.
  \label{EQsigmaDynamicalAsPoissonBracket}
\end{align}
which mirrors the Poisson bracket form of the static shifting
operators $\bsig(\rv)$ given in \eqr{EQbsigAsPoissonBracket}.  The
equivalence of Eqs.~\eqref{EQsigmaDynamicalAsPropagatorSandwich} and
\eqref{EQsigmaDynamicalAsPoissonBracket} is proven by starting from
the right hand side of the latter and re-writing as follows:
$\{\,\cdot\,, m\hat\Jv(\rv,t)\} = \{\calU^\dagger(t,0)\calU(t,0)
\,\cdot\,, \calU^\dagger(t,0) m\hat \Jv(\rv)\} =
\calU^\dagger(t,0)\{\calU(t,0)\,\cdot\,,m\hat\Jv(\rv)\} =
\calU^\dagger(t,0)\{\,\cdot\,,m\hat\Jv(\rv)\}\calU(t,0) =
\calU^\dagger(t,0)\bsig(\rv)\calU(t,0)$, where we have first inserted
an identity operator $\calU^\dagger(t,0)\calU(t,0)=1$ and then
exploited that the propagator and the Poisson bracket commute, due to
the latter being a canonical invariant under the Hamiltonian time
evolution.

Besides their anti-self-adjointness
\eqref{EQbsigDynamicalAntiSelfAdjoint}, the dynamical shifting
operators $\bsig(\rv,t)$, as expressed in the alternative form
\eqref{EQsigmaDynamicalAsPoissonBracket}, possess two further key
properties.  First, when applied to the (thermally scaled negative)
Hamiltonian, one obtains
\begin{align}
  \hat\Cv(\rv,t) &= -\bsig(\rv,t)\beta H_0.
  \label{EQshiftCurrentFromBsig}
\end{align}
where $\hat \Cv(\rv,t)$ is the shift current phase space function
defined via \eqr{EQdefinitionOfBare}. The validity of
\eqr{EQshiftCurrentFromBsig} can be seen by expressing its right hand
side via the Poisson bracket \eqref{EQsigmaDynamicalAsPoissonBracket}
as $-\{\beta H_0, m\hat\Jv(\rv,t)\} = \beta \{m\hat\Jv(\rv,t),
H_0\}=\beta L_0 m\hat\Jv(\rv,t)$, where identifying the initial state
Liouvillian $L_0$ gives the right hand side of
\eqr{EQdefinitionOfBare}.  Secondly, when applying $\bsig(\rv,t)$ to
the initial state probability distribution $f_0$, given by the
Boltzmann form~\eqref{EQf0}, then
\begin{align}
  \bsig(\rv,t) f_0 &= \hat\Cv(\rv,t) f_0,
  \label{EQeigenvalueProperty}
\end{align}
as follows from the chain rule and identifying $\hat \Cv(\rv,t)$ via
\eqr{EQshiftCurrentFromBsig}.

As a demonstration of the power of the present dynamical shifting
operator formalism, we first reconsider the derivation of the shift
current sum rule \eqref{EQbareHyperForceBalance}.  Viewing this
identity from the perspective of the dynamical shifting gauge
invariance allows one to formulate a strictly valid derivation in the
following strikingly compact form: $\Cv(\rv,t)=\Tr \hat \Cv(\rv,t) f_0
= \Tr \bsig(\rv,t)f_0 = -\Tr f_0\bsig(\rv,t)1 = 0$, where we have
first spelled out the average $\langle \hat\Cv(\rv,t)\rangle$, then
introduced $\bsig(\rv,t)$ via \eqr{EQeigenvalueProperty}, and
reordered terms by using that $\bsig(\rv,t)$ is anti-self-adjoint
\eqref{EQbsigDynamicalAntiSelfAdjoint}. That the resulting average
vanishes is trivially due to $\bsig(\rv,t)1=0$, as follows from
\eqr{EQsigmaDynamicalAsPoissonBracket} and the vanishing Poisson
bracket of any constant.

The connection of the dynamical generalization
\eqref{EQsigmaDynamicalAsPoissonBracket} of equilibrium phase space
shifting \cite{mueller2024gauge, mueller2024whygauge} with the
nonequilibrium shift current sum rule \eqref{EQbareHyperForceBalance}
reveals the latter to originate from dynamical gauge invariance. As we
demonstrate in the following, the dynamical gauge concept allows one
to address a significantly wider class of cases than covered thus far,
by incorporating general hyperobservables into the framework.

\subsection{Exact hypercurrent sum rules for general observables}
\label{SEChyperobersableSumRules}

We consider a general phase space function $\hat A(\rv^N,\pv^N)$ to
act as a hyperobservable that is of physical interest in the
nonequilibrium situation under consideration. We recall both that the
corresponding Heisenberg observable $\hat A(\rv^N, \pv^N, t)$ is
obtained via application of the adjoint propagator according to $\hat
A(t) = \calU^\dagger(t,0)\hat A$, see Sec.~\ref{SECtimeEvolution}, and
that trajectory-based methods facilitate alternative access, as
described in Sec.~\ref{SECtrajectories}.  We follow
Refs.~\cite{robitschko2024any, mueller2024gauge, mueller2024whygauge}
in defining the {\it static} hyperforce density as the phase space
function given by
\begin{align}
  \hat \Sv_A(\rv) &= \bsig(\rv) \hat A,
  \label{EQhyperForceOperatorStatic}
\end{align}
where we recall the explicit form \eqref{EQbsigExplicit} of the static
differential operators $\bsig(\rv)$. The corresponding Heisenberg
observable $\hat\Sv_A(\rv,t)$ then follows via the standard procedure
as $\hat \Sv_A(\rv,t)=\calU^\dagger(t,0)\hat \Sv_A(\rv)$. Using
\eqr{EQhyperForceOperatorStatic} and inserting an identity operator
$\calU(t,0)\calU^\dagger(t,0)=1$ leads to
$\hat\Sv_A(\rv,t)=\calU^\dagger(t,0)\bsig(\rv)\calU(t,0)\calU^\dagger(t,0)\hat
A$.  The first three factors therein can be identified as the
dynamical shifting operator $\bsig(\rv,t)$, see
\eqr{EQsigmaDynamicalAsPropagatorSandwich}, and the remaining two
factors constitute the Heisenberg observable $\hat A(t)$, see
\eqr{EQhatAdynamicalFromU}. Hence we formulate the {\it dynamical}
hyperforce density as the following Heisenberg observable:
\begin{align}
  \hat\Sv_A(\rv,t) &= \bsig(\rv,t) \hat A(t),
  \label{EQhyperForceOperatorDynamic}
\end{align}
which generalizes the static equivalent $\hat\Sv_A(\rv)$ given by
\eqr{EQhyperForceOperatorStatic}.  Building the dynamical average in
the standard way gives the dynamical hyperforce density as
$\Sv_A(\rv,t) =\langle \hat \Sv_A(\rv,t)\rangle$. As a complement, in
trajectory-based work, for a given form of $\hat\Sv_A(\rv)$,
straightforward evaluation gives $\hat\Sv_A(\rv,t) = \hat
\Sv_A(\rv;\hat\rv^N(t), \hat\pv^N(t))$, where we have written out the
phase space dependence to emphasize that this expression is ready to
be averaged over initial states once the trajectory is known for the
considered initial microstates; recall that the static hyperforce
density $\hat\Sv_A(\rv;\rv^N,\pv^N) = \bsig(\rv) \hat A$ is explicitly
available for given form of the hyperobservable $\hat A$
\cite{robitschko2024any, mueller2024gauge, mueller2024whygauge}.

Spelling out the average allows one to proceed in explicit form:
$\Sv_A(\rv,t)=\Tr f_0 \bsig(\rv,t) \hat A(t) = \Tr \hat A(t)
\bsig^\dagger(\rv,t) f_0$, where we have introduced the shifting
operator via \eqr{EQhyperForceOperatorDynamic} and then reordered the
integrand via building the adjoint. Using the anti-self-adjoint
property \eqref{EQbsigDynamicalAntiSelfAdjoint} of $\bsig(\rv,t)$
allows one to rewrite the result as $-\Tr \hat A(t) \bsig(\rv,t) f_0 =
-\Tr \hat A(t) \hat\Cv(\rv,t) f_0 =-\langle \hat A(t) \hat \Cv(\rv,t)
\rangle =-\langle \hat \Cv(\rv,t) \hat A(t) \rangle$, where we have
used the generation \eqref{EQeigenvalueProperty} of the shift current
Heisenberg observable $\hat\Cv(\rv,t)$, then have written the overall
result as a nonequilibrium average, and in the last step have
interchanged the two phase space functions.  Bringing the result to
the initial left hand side yields the following dynamical hypercurrent
balance:
\begin{align}
  \Sv_A(\rv,t) + \langle  \hat \Cv(\rv,t) \hat A(t) \rangle
  &= 0,
  \label{EQdressedHyperForceBalance}
\end{align}
which is exact. Using compact notation for the hypercurrent
correlation function, $\Cv_A(\rv,t) = \langle \hat\Cv(\rv,t) \hat A(t)
\rangle$, allows one to express the sum rule
\eqref{EQdressedHyperForceBalance} in the alternative form:
\begin{align}
  \Sv_A(\rv,t) + \Cv_A(\rv,t) &= 0,
\end{align}
which expresses the vanishing sum of hyperforce density (first term)
and hypercurrent correlation function (second term).

As a simple consistency check, making the choice $\hat A=1$ leads to
$\hat\Sv_{\hat A=1}(\rv,t)=0$ and hence on average $\Sv_{\hat
  A=1}(\rv,t)=0$. Thus upon recalling $\langle
\hat\Cv(\rv,t)\rangle=\Cv(\rv,t)=\Cv_{\hat A=1}(\rv,t)$, the
hypercurrent identity~\eqref{EQdressedHyperForceBalance} reduces to
$\Cv(\rv,t)=0$, which constitutes the shift current identity
\eqref{EQbareHyperForceBalance}. As an aside, the hypercurrent
correlation term in \eqr{EQdressedHyperForceBalance} can alternatively
be expressed as $\langle \hat \Cv(\rv,t) \hat A(t) \rangle = \cov(\hat
\Cv(\rv,t), \hat A(t))$, where the covariance of two observables is
defined in the standard form: $\cov(\hat \Cv(\rv,t), \hat A(t))=
\langle \hat\Cv(\rv,t) \hat A(t) \rangle - \Cv(\rv,t) \langle \hat
A(t)\rangle$. The equivalence of correlation and covariance follows
from $\Cv(\rv,t)=0$, see the shift current
balance~\eqref{EQbareHyperForceBalance}.  Working with covariances can
have practical advantages in simulation work due to systematic
subtraction of sampling uncertainties, see
e.g.\ Ref.~\cite{robitschko2024any} for such work in equilibrium.

The dynamical correlation function $\langle \hat\Cv(\rv,t) \hat A(t)
\rangle$ that features in \eqr{EQdressedHyperForceBalance} can be
split into ideal, interparticle, and external contributions, according
to $ \langle \hat \Cv(\rv,t) \hat A(t) \rangle = \langle
\hat\Cv_\rmid(\rv,t) \hat A(t) \rangle +\langle \hat\Cv_\rmint(\rv,t)
\hat A(t) \rangle + \langle \hat\Cv_\rmext(\rv,t) \hat A(t) \rangle,$
where we recall the definition of the ideal, interparticle, and
external particle shift current, as arising from the splitting of the
initial state Liouvillian; see the corresponding description given
below \eqr{EQbareHyperForceBalance}. The splitting allows one to
express the hypercurrent sum rule~\eqref{EQdressedHyperForceBalance}
in equivalent form as:
\begin{align}
  & \Sv_A(\rv,t) + \Cv_A^\rmid(\rv,t) + \Cv_A^\rmint(\rv,t)
  +\Cv_A^\rmext(\rv,t)  = 0,
  \label{EQdressedHyperForceBalanceSplitForm}
\end{align}
where the three partial hypercurrent correlation functions are defined
as $\Cv_A^\rmid(\rv,t)=\langle \hat\Cv_\rmid(\rv,t) \hat A(t)
\rangle$, $\Cv_A^\rmint(\rv,t)=\langle \hat\Cv_\rmint(\rv,t) \hat A(t)
\rangle$, and $\Cv_A^\rmext(\rv,t)= \langle \hat\Cv_\rmext(\rv,t) \hat
A(t) \rangle$.  We defer a detailed description of the connection of
the dynamical identity \eqref{EQdressedHyperForceBalanceSplitForm}
with the static hyperforce theory \cite{robitschko2024any,
  mueller2024gauge, mueller2024whygauge} to
Sec.~\ref{SEClinkToEquilibrium} below.

The present hypercurrent formalism rests centrally on the properties
of the dynamical shifting operators $\bsig(\rv,t)$, as alternatively
given by Eqs.~\eqref{EQsigmaDynamicalAsPropagatorSandwich} and
\eqref{EQsigmaDynamicalAsPoissonBracket}. The application of these
differential operators generates both the dynamical hyperforce density
$\hat\Sv_A(\rv,t)$, given by \eqr{EQhyperForceOperatorDynamic}, and
the dynamical shift current $\hat\Cv(\rv,t)$, which arises from
\eqr{EQshiftCurrentFromBsig}. As $\hat\Cv(\rv,t)$ is also given as the
phase space function~\eqref{EQdefinitionOfBare}, the emerging temporal
correlation structure rests centrally on the relevance of the initial
state Liouvillian $L_0$ for the {\it nonequilibrium} problem, as we
expand on in the following.

\subsection{Initial state time differentiation}
\label{SECinitialStateTimeDifferentiation}

Identifying and applying the initial state Liouvillian~$L_0$ to act
inside of a nonequilibrium average forms a crucial step in the above
laid out hypercurrent theory, see the central definition
\eqref{EQdefinitionOfBare} of $\hat \Cv(\rv,t)$. We here explore the
implied temporal structure further. For a given Heisenberg
hyperobservable $\hat A(\rv^N, \pv^N, t)$ we define its {\it initial
  state time derivative} as
\begin{align}
  \hat A^\mystar(t) 
  &= L_0 \hat A(t)
  \label{EQAhatBullet}\\
  & =  \frac{\partial}{\partial s}\calU_0^\dagger(s,0) \hat A(t)
  \Big|_{s=0},
  \label{EQAhatBulletAlternative}
\end{align}
where the superscript bold dot denotes the application of $L_0$ to
$\hat A(t)$ in \eqr{EQAhatBullet} such that spelling out the initial
state Liouvillian in Poisson bracket form \eqref{EQLiouvillian}
implies $\hat A^\bullet(t)=\{\hat A(t), H_0\}$. In the alternative
form \eqref{EQAhatBulletAlternative} the adjoint initial state
propagator $\calU_0^\dagger(s,0)=\e^{sL_0}$ performs the initial state
dynamics from time 0 to time $s$.  We recall that the initial
ensemble, as characterized by the Hamiltonian $H_0$ and corresponding
thermal distribution $f_0$ given by \eqr{EQf0}, remains stationary
under its inherent time evolution according to its corresponding
Liouvillian $L_0$; see the description of this setup in
Sec.~\ref{SECnonequilibriumDynamics}.

To make the connection of the initial state time differentitation
\eqref{EQAhatBullet} with the shift current observable $\hat
\Cv(\rv,t)$, as given by \eqr{EQdefinitionOfBare}, we first define the
scaled one-body current $\hat\Sv(\rv,t)=\beta m \hat\Jv(\rv,t)$, where
we recall the definition~\eqref{EQcurrentOperator} of the one-body
current and that $\beta$ is the inverse temperature of the initial
state. The shift current is then obtained by applying the initial
state Liouvillian $L_0$ according to \eqr{EQAhatBullet}, which yields
\begin{align}
  \hat\Cv(\rv,t) &=  \hat\Sv^\bullet(\rv,t) = L_0 \hat\Sv(\rv,t).
  \label{EQSvBullet}
\end{align}
such that $\hat\Cv(\rv,t) = \beta m \hat\Jv^\bullet(\rv,t)$, as
follows from \eqr{EQdefinitionOfBare}. As an aside, the explicit
dependence on mass $m$ scales out, as described in
Sec.~\ref{SECdynamicalHyperforceCorrelations} and further elaborated
in Sec.~\ref{SECmolecularDynamicsConcepts} below.

Using the bold dot notation \eqref{EQAhatBullet}, the hypercurrent sum
rule \eqref{EQdressedHyperForceBalance} attains the following form:
\begin{align}
  \Sv_A(\rv,t) + \langle \hat\Sv^\bullet(\rv,t) \hat A(t) \rangle 
  &= 0.
\end{align}
We next derive a further nonequilibrium sum rule that is satisfied by
the second term on the above left hand side, which is the hypercurrent
correlation function $\Cv_A(\rv,t)=\langle \hat\Sv^\bullet(\rv,t) \hat
A(t) \rangle$.  Writing out explicitly the implied average allows one
to proceed as follows: $\Tr f_0 \hat A(t) \hat\Sv^\mystar(\rv,t) = \Tr
f_0\hat A(t)L_0 \hat \Sv(\rv,t) = \Tr \hat\Sv(\rv,t)L_0^\dagger\hat
A(t) f_0$, where we have first used \eqr{EQSvBullet} to reintroduce
$L_0$ and then reordered via building the adjoint. We next use that
the initial state Liouvillian is anti-self-adjoint, $L_0^\dagger =
-L_0$, and that the initial state distribution $f_0$ is stationary
under the initial time evolution, $L_0 f_0=0$. Upon reordering the
result implies the following exact hypercurrent `swap' identity:
\begin{align}
  \langle \hat\Bare^\mystar(\rv,t)  \hat A(t) \rangle
  + \langle  \hat\Bare(\rv,t) \hat A^\mystar(t)  \rangle &= 0,
  \label{EQdressedHyperCorrelatorAlternative}
\end{align}
where we recall that according to the notation \eqref{EQAhatBullet},
$\hat A^\mystar(t)=L_0\hat A(t) = \{\hat A(t), H_0\}$ is the initial
state derivative of the Heisenberg observable $\hat A(t)$.

We can combine Eqs.~\eqref{EQdressedHyperForceBalance} and
\eqref{EQdressedHyperCorrelatorAlternative} to obtain the following
alternative hypercurrent sum rule:
\begin{align}
  \Sv_A(\rv,t)  - \langle \hat\Bare(\rv,t) \hat A^\mystar(t)  \rangle &= 0.
\end{align}

The theory thus far developed applies to general observables $\hat A$
and it is formally exact for general nonequilibrium Hamiltonian
dynamics, as generated by a time-dependent many-body Hamiltonian $H$
and starting from an initial thermal state with Hamiltonian $H_0$ and
inverse temperature $\beta$. We next describe the implications when
working with trajectories.

\subsection{Trajectory-based differentiation}
\label{SECmolecularDynamicsConcepts}

As laid out in Sec.~\ref{SECtrajectories} a trajectory-based picture
rests on having access to $\hat\rv_i(t)$ and $\hat\pv_i(t)$ for all
particles $i=1,\ldots,N$. We recall that the full phase space
dependence, when viewing the particle positions and momenta as
Heisenberg observables, is $\hat\rv_i(\rv^N, \pv^N,t)$ and
$\hat\pv_i(\rv^N, \pv^N,t)$, where the argument $\rv^N, \pv^N$ denotes
the phase space point that represents the initial configuration at
$t=0$. One can express this relationship succinctly at the initial
time as: $\hat\rv_i(\rv^N,\pv^N,0)=\rv_i$ and
$\hat\pv_i(\rv^N,\pv^N,0)=\pv_i$.

We hence initial-state time differentiate the entire trajectory as
follows:
\begin{align}
  \hat\rv_i^\mystar(t) &= L_0 \hat\rv_i(t) = \{\hat\rv_i(t), H_0\}
  \notag\\
  &=\sum_j \Big(\vel_{j,0}\cdot\nabla_j \hat\rv_i(t)
  + \fv_{j,0}\cdot \nabla_{\pv_j} \hat \rv_i(t)
  \Big),
  \label{EQrviBullet}
  \\
  \hat\pv_i^\mystar(t) &= L_0 \hat\pv_i(t) = \{\hat\pv_i(t), H_0\}
  \notag\\
  &=\sum_j \Big(\vel_{j,0}\cdot\nabla_j \hat\pv_i(t)
  + \fv_{j,0}\cdot \nabla_{\pv_j} \hat \pv_i(t)
  \Big),
  \label{EQpviBullet}
\end{align}
where $\vel_{j,0}=\pv_j/m_0$ denotes the initial state velocity,
$\fv_{j,0}=-\nabla_j H_0$ is the initial state force of particle $j$,
and the sums run over all particles $j=1,\ldots, N$.  The phase space
derivatives $\nabla_j$ and $\nabla_{\pv_j}$ can be viewed as measuring
the dependence of position $\hat\rv_i(t)$ in \eqr{EQrviBullet} and of
momentum $\hat\pv_i(t)$ in \eqr{EQpviBullet} of particle $i$ upon
changes in the initial data of particle $j$. Rather than forming a
generic perturbation, the phase space derivatives are then weighted
specifically by, respectively, the initial state velocity and initial
force of particle $j$. Summation over all particles $j$ then yields
$\hat\rv_i^\bullet(t)$ and $\hat\pv_i^\bullet(t)$. In relation to the
Liouvillian splitting \eqref{EQLsplitting} into ideal, interparticle,
and external contributions, the first term in the sum in both
Eqs.~\eqref{EQrviBullet} and \eqref{EQpviBullet} arises from $L_\rmid$
and the second term arises from the sum $L_\rmint + L_\rmext$.

The initial state derivative of a general Heisenberg observable $\hat
A(t)$ then follows from using the chain rule, which gives
\begin{align}
  \hat A^\mystar(t) &= 
  \sum_i \Big[
    (\nabla_i \hat A)(t) \cdot \hat\rv_i^\mystar(t)
    +  (\nabla_{\pv_i} \hat A)(t) \cdot \hat\pv_i^\mystar(t)
    \Big],
\end{align}
where we use the notation $(\nabla_i\hat
A)(t)=\calU^\dagger(t,0)\nabla_i \hat A$ and $(\nabla_{\pv_i}\hat
A)(t)=\calU^\dagger(t,0)\nabla_{\pv_i} \hat A$ to respectively
indicate the temporally propagated phase space position and momentum
gradients of the hyperobservable $\hat A(\rv^N,\pv^N)$.

We next turn to the shift current \eqref{EQSvBullet}, which we recall
as $\hat\Cv(\rv,t) = \hat\Sv^\mystar(\rv,t)$ with $\hat\Sv(\rv,t) =
\beta m \hat\Jv(\rv,t)$. The present formulation allows one to obtain
the following trajectory-based representation: $\hat \Cv(\rv,t) =
\sum_i [\hat\rv_i^\mystar(t) \cdot (\nabla_i \hat \Sv)(\rv,t)
  +\hat\pv_i^\mystar(t) \cdot (\nabla_{\pv_i} \hat \Sv)(\rv,t)]$. We
express the sum of these two terms as
\begin{align}
  \hat\Cv(\rv,t) &= \nabla \cdot
  \hat\taub_C(\rv,t) + \hat\Cv_\rmacc(\rv,t),
  \label{EQCvSplittingAccelerationTransport}
\end{align}
where the shift stress tensor $\hat\taub_C(\rv,t)$ and the shift
acceleration current observable $\hat\Cv_\rmacc(\rv,t)$ are given
respectively by
\begin{align}
  \hat\taub_C(\rv,t) &=
  -\beta \sum_i\delta\big(\rv-\hat\rv_i(t)\big)
  \hat\rv_i^\bullet(t)\hat\pv_i(t),
  \label{EQtauChatDefinition}\\
  \hat\Cv_\rmacc(\rv,t) &=
  \beta\sum_i \delta\big(\rv-\hat\rv_i(t)\big)\hat\pv_i^\bullet(t).
  \label{EQCacchatDefinition}
\end{align}

That the shift current $\hat\Cv(\rv,t)$ has the natural splitting
\eqref{EQCvSplittingAccelerationTransport} mirrors the decomposition
of the force density $\hat\Fv(\rv,t)$, given below
\eqr{EQforceDensityOperator} in the form: $\hat\Fv(\rv,t) =
\nabla\cdot \hat\taub(\rv,t) + \hat \Fv_U(\rv,t)$, where the potential
force density observable, $\hat\Fv_U(\rv,t)=\hat\Fv_\rmint(\rv,t) +
\hat\Fv_\rmext(\rv,t)$, combines interparticle and external
contributions.
Note also the structural similarity of \eqr{EQtauChatDefinition} with
the definition of the standard kinetic stress tensor observable
$\hat\taub(\rv,t)$, given in Schr\"odinger form below
\eqr{EQforceDensityOperator}, and of the shift acceleration current
observable \eqref{EQCacchatDefinition} with $\hat \Fv_U(\rv,t)$.

Upon averaging, one can show that the mean shift stress tensor, as
given by $\taub_C(\rv,t) =\langle \hat\taub_C(\rv,t) \rangle$, and the
mean shift acceleration current, as given by $\Cv_\rmacc(\rv,t) =
\langle \hat\Cv_\rmacc(\rv,t) \rangle$, are related to the {\it
  instantaneous} dynamical density profile $\rho(\rv,t)$ and its
gradient respectively via:
\begin{align}
  \taub_C(\rv,t) &=  -\rho(\rv,t) \unity
  \label{EQtauCsumrule}\\
  \Cv_\rmacc(\rv,t)  &= \nabla \rho(\rv,t),
  \label{EQCaccIdentity}
\end{align}
where $\unity$ denotes the $d\times d$-unit matrix.  The identity
\eqref{EQCaccIdentity} follows straightforwardly from writing out the
average on the left hand side explicitly; we give a compact account:
$\Cv_\rmacc(\rv,t)=\Tr f_0 \hat \Cv_\rmacc(\rv,t) = \Tr f(t)
\calU(t,0)\sum_i \delta(\rv-\hat\rv_i(t))\{\hat\pv_i(t), \beta H_0\}$.
Then applying the propagator and subsequently evaluating the Poisson
bracket gives:
$\Tr f(t) \sum_i \delta(\rv-\rv_i)\{\pv_i, \calU(t,0) \beta H_0\} =
-\Tr f(t) \sum_i \delta(\rv-\rv_i) \nabla_i [\calU(t,0)\beta H_0] =
\Tr \sum_i\delta(\rv-\rv_i)\nabla_i f(t) = \nabla\rho(\rv,t)$, where
we have used that $f(t) = \e^{-\beta \calU(t,0) H_0}/Z_0$ and the last
step follows from integration by parts with respect to the particle
positions, with $\nabla_i\delta(\rv-\rv_i)=-\nabla\delta(\rv-\rv_i)$
and then identifying the average as the gradient of the dynamical
density profile. Similar steps prove \eqr{EQtauCsumrule}.

The identities \eqref{EQtauCsumrule} and \eqref{EQCaccIdentity} verify
explicitly the shift current sum rule \eqref{EQbareHyperForceBalance}
in the form
\begin{align}
  \Cv_\rmacc(\rv,t)+\nabla\cdot \taub_C(\rv,t)=0.
  \label{EQbareHyperForceBalanceSplitForm}
\end{align}
Applying the splitting \eqref{EQCvSplittingAccelerationTransport} into
transport and acceleration contributions analogously to the
hypercurrent correlation function $\Cv_A(\rv,t)=\langle
\hat\Cv(\rv,t)\hat A(t)\rangle$, as it appears in the hypercurrent sum
rule \eqref{EQdressedHyperForceBalance}, leads to the following
decomposition:
\begin{align}
  \Cv_A(\rv,t) &=
  \langle \hat \Cv_\rmacc(\rv,t) \hat A(t) \rangle + \nabla \cdot \langle
  \hat\taub_C(\rv,t) \hat A(t) \rangle,
  \label{EQCvAsplitting}
\end{align}
where each term on the right had side can further be split into ideal,
interparticle, and external contributions using the Liouvillian
splitting \eqref{EQLsplitting} for the initial state derivative.

For completeness, in explicit form the classical hyperforce density
observable $\hat \Sv_A(\rv)$, see
Eqs.~\eqref{EQhyperForceOperatorStatic} for its emergence from
applying the shifting operator to the observable~$\hat A$, is given by
\begin{align}
  \hat \Sv_A(\rv) &= \sum_i \delta(\rv-\rv_i)\nabla_i \hat A
  + \nabla \cdot \sum_i \delta(\rv-\rv_i)(\nabla_{\pv_i}\hat A) \pv_i.
  \label{EQJohanna0}
\end{align}
One can decompose as $\hat\Sv_A(\rv) = \hat\Sv_A^{\rm pos}(\rv) +
\nabla\cdot \hat\taub_A(\rv)$, where the position contribution to the
hyperforce density is $\hat\Sv_A^{\rm pos}(\rv)= \sum_i
\delta(\rv-\rv_i)\nabla_i \hat A$ and the hyperstress tensor is given
by $\hat\taub_A(\rv)= \sum_i \delta(\rv-\rv_i)(\nabla_{\pv_i}\hat
A)\pv_i$.

Following arguments that are very similar to those given to derive
\eqr{EQCaccIdentity}, one can further show that
\begin{align}
  \langle \hat\taub_C(\rv,t) \hat A(t)  \rangle  + \taub_A(\rv,t)
  &= - \langle \hat \rho(\rv,t) \hat A(t) \rangle \unity,
  \label{EQJohanna2}
  \\
  \langle \hat \Cv_\rmacc(\rv,t) \hat A(t) \rangle
  + \Sv_A^{\rm pos}(\rv,t)
  &= \nabla \langle \hat\rho(\rv,t) \hat A(t) \rangle,
  \label{EQJohanna1}
\end{align}
where $\taub_A(\rv,t) = \langle \hat\taub_A(\rv,t) \rangle$ and
$\Sv_A^{\rm pos}(\rv,t) = \langle \hat \Sv_A^{\rm pos}(\rv,t)
\rangle$.  As a consistency check, adding \eqr{EQJohanna1} and the
divergence of \eqr{EQJohanna2} yields the dynamical hypercurrent sum
rule \eqref{EQdressedHyperForceBalance}, when identifying the
hypercurrent correlation function $\Cv_A(\rv,t)$ via
\eqr{EQCvAsplitting} and the hyperforce density as $\Sv_A(\rv,t)=
\Sv_A^{\rm pos}(\rv,t) + \nabla\cdot\taub_A(\rv,t)$, as follows from
the splitting given below \eqr{EQJohanna0}.

\subsection{Concrete choices of hyperobservables}
\label{SECconcreteSumRules}

To illustrate the broad range of consequences of the above laid out
dynamical statistical mechanical gauge invariance, we specialize the
general hypercurrent sum rule \eqref{EQdressedHyperForceBalance} for
several concrete choices for the hyperobservable~$\hat A$. We first
consider both the Hamiltonian, $\hat A=H$, and the interparticle
potential, $\hat A=u(\rv^N)$, which respectively yields:
\begin{align}
  \Fv(\rv,t) 
   -\langle \hat\Cv(\rv,t) H(t)\rangle &= 0,
   \label{EQdressedIdentity1}\\
   \Fv_\rmint(\rv,t) 
   -\langle \hat\Cv(\rv,t) u(t)  \rangle &= 0,
   \label{EQdressedIdentity2}
\end{align}
where we recall $\Fv(\rv,t)$ as the total dynamical mean force density
and $\Fv_\rmint(\rv,t)$ as its interparticle contribution; see their
descriptions below \eqr{EQforceDensityOperator}.

As two further specific choices, we take the total momentum, $\hat A=
\hat \Pv= \sum_i\pv_i$, and the sum of positions, $\hat A= \hat\Rv =
\sum_i\rv_i$, where the latter can be viewed as the product of the
center of mass, $\hat \Rv/N$, and the total particle number $N$. The
generic hypercurrent sum rule \eqref{EQdressedHyperForceBalance} then
attains the following two respective forms:
\begin{align}
  \nabla m\Jv(\rv,t)
   + \langle \hat \Cv(\rv,t) \hat\Pv(t)  \rangle &= 0,
   \label{EQdressedIdentity3}\\
   \rho(\rv,t)\unity 
   + \langle \hat \Cv(\rv,t) \hat\Rv(t) \rangle &= 0.
   \label{EQdressedIdentity4}
\end{align}
The four sum rules
\eqref{EQdressedIdentity1}--\eqref{EQdressedIdentity4} are exact and
they are noteworthy, as in each instance a very common instantaneous
observable (first term on each left hand side) is related rigorously
to a specific correlation function (corresponding second term) with
the shift current.  Recalling the initial-time
form~\eqref{EQdefinitionOfBare} of the shift current operator $\hat
\Cv(\rv,t)$ reveals the inherent temporally nonlocal character of
these correlation functions; in contrast the corresponding hyperforce
density [as realized by the respective first term in
  \eqref{EQdressedIdentity1}--\eqref{EQdressedIdentity4}] carries
instantaneous dependence on time $t$.

On the basis of the swap identity
\eqref{EQdressedHyperCorrelatorAlternative} we can alternatively
formulate each of the second terms in
Eqs.~\eqref{EQdressedIdentity1}--\eqref{EQdressedIdentity4} by using
the scaled current $\hat \Sv(\rv,t)=m\beta \hat\Jv(\rv,t)$ instead of
the shift current $\hat\Cv(\rv,t)=\Sv^\mystar(\rv,t)=L_0\beta m
\hat\Jv(\rv,t)$. We obtain explicitly from
\eqr{EQdressedHyperCorrelatorAlternative} for the Hamiltonian
correlation function in \eqr{EQdressedIdentity1}: $\langle \hat
\Cv(\rv,t) H(t) \rangle = -\langle \hat \Sv(\rv,t) H^\mystar(t)
\rangle,$ for the interparticle potential in \eqr{EQdressedIdentity2}:
$\langle \hat \Cv(\rv,t) u(t) \rangle = -\langle \hat \Sv(\rv,t)
u^\mystar(t) \rangle,$ for the total momentum in
\eqr{EQdressedIdentity3}: $\langle \hat \Cv(\rv,t) \hat\Pv(t) \rangle
= -\langle \hat\Sv(\rv,t) \hat\Pv^\mystar(t) \rangle$, and for the sum
of particle positions in \eqr{EQdressedIdentity4}: $\langle \hat\Rv(t)
\hat \Cv(\rv,t) \rangle = -\langle \hat\Rv^\mystar(t) \hat\Sv(\rv,t)
\rangle$, where we recall that the superscript bold dot indicates the
application of the initial state Liouvillian $L_0$ according to
\eqr{EQAhatBullet}.

\section{Applications}
\label{SECapplications}

\subsection{Equilibrium limit of hypercurrent correlations}
\label{SEClinkToEquilibrium}

We first demonstrate the consistency of the dynamical gauge theory
formulated in Sec.~\ref{SECdynamicalGaugeInvariance} with the static
hyperforce approach of Refs.~\cite{robitschko2024any,mueller2024gauge,
  mueller2024whygauge}. We hence restrict the above general
nonequilibrium setup, as described in detail in
Sec.~\ref{SECnonequilibriumDynamics}, to cases with no switching at
the initial time, such that the Hamiltonian remains unchanged,
$H=H_0$, at all times.  We consider the implications for the
hypercurrent sum rule~\eqref{EQdressedHyperForceBalance} which we
recall contains for the special case $\hat A=1$ the shift current
balance \eqref{EQshiftingCurrentVanishes}. The first term on the left
hand side of \eqr{EQdressedHyperForceBalance}, i.e., the dynamical
hyperforce density $\Sv_A(\rv,t)$, simply becomes $\Sv_A(\rv)$,
independent of time. The hypercurrent correlation function is $\langle
\hat\Cv_0(\rv,t) \hat A(t)\rangle$, where the subscript 0 denotes the
equilibrium shift current observable $\hat\Cv_0(\rv,t) = \beta L_0 m_0
\hat \Jv_0(\rv,t)$ as is obtained from \eqr{EQdefinitionOfBare} for
the initial state. We can re-write this on the basis of
Eq.~\eqref{EQdefinitionOfBare} as: $\beta L_0 \calU_0^\dagger(t,0)
m_0\hat \Jv_0(\rv) = \beta \calU_0^\dagger(t,0) L_0 m_0 \hat\Jv_0(\rv)
= \beta \calU_0^\dagger(t,0) \hat\Fv_0(\rv) = \beta \hat\Fv_0(\rv,t)$,
where we have first exploited that $L_0$ and $\calU_0^\dagger(t,0)$
commute and then have identified the initial state force density
observable $\hat\Fv_0(\rv,t)$, as corresponds to $H_0$; see
Sec.~\ref{SEConeBodyLevel}.

Overall we hence obtain $\langle \hat\Cv_0(\rv,t) \hat A(t) \rangle =
\langle \beta \hat\Fv_0(\rv,t) \hat A(t) \rangle = \langle
\beta\hat\Fv_0(\rv) \hat A\rangle$, where the time dependence in the
last step vanishes due to stationarity in equilibrium.  We hence
obtain the following equilibrium identity:
\begin{align}
  \Sv_A(\rv) + \langle\beta \hat\Fv_0(\rv)  \hat A \rangle=0,
  \label{EQhyperForceBalanceStatic}
\end{align}
which is the static hyperforce balance \cite{robitschko2024any,
  mueller2024gauge, mueller2024whygauge}.  The force correlation
function can thereby be decomposed into its three constituent (ideal,
interparticle, and external) parts, such that alternatively to
\eqr{EQhyperForceBalanceStatic} we can write
\begin{align}
  & \Sv_A(\rv) + \langle  \beta \hat\Fv_{\rmid,0}(\rv) \hat A\rangle +
   \langle \beta \hat\Fv_{\rmint,0}(\rv) \hat A \rangle
   \notag\\&\quad\quad\;\;
   +\langle \beta \hat\Fv_{\rmext,0}(\rv) \hat A \rangle  =0,
\end{align}
where the individual force density contributions, given in general
form below \eqr{EQcurrentOperator}, are those for the initial
Hamiltonian $H_0$.

The above reduction to the static gauge invariance hyperforce theory
constitutes an important consistency check. However, much more
interesting structure is revealed by not enforcing the static limit
but rather honouring the equilibrium {\it dynamics}. As before, we
consider the situation of continuing equilibrium, such that $H=H_0$ at
all times with no switching nor any further explicit time dependence
occurring in $H$. As described above, the hyperforce density then
looses its time dependence, $\Sv_A(\rv,t)=\Sv_A(\rv)$. However, such
reduction does not occur in general when splitting the hypercurrent
correlation function, $\langle \hat \Cv(\rv,t) \hat A(t) \rangle$, as
we demonstrate in the following.

We hence forgo the operator re-ordering that led to
\eqr{EQhyperForceBalanceStatic} and retain the structure
\eqref{EQdefinitionOfBare} of the shift current observable:
$\hat\Cv(\rv,t) = \beta L_0 m\hat\Jv(\rv,t)$. Then the present
equilibrium situation allows one to obtain from splitting the
Liouvillian the result: $\hat \Cv_{0,\alpha}(\rv,t) = \beta
L_{0,\alpha} m_0 \Jv(\rv,t)$, where the subscript $\alpha=$ `$\rmid$',
`$\rmint$', and `$\rmext$' labels the different ideal, interparticle,
and external contributions.  In contrast to the above derivation of
the static hyperforce sum rule \eqref{EQhyperForceBalanceStatic}, here
no further simplification arises, as the partial Liouvillians
$L_{0,\alpha}$ do {\it not} in general commute with the full initial
state adjoint propagator $\calU_0^\dagger(t,0)$.

We can conclude that two properties render the present dynamical
equilibrium limit non-trivial and different from the static case: i)
In general the partial hyperforce and hypercurrent correlation
functions differ from each other, $\langle \beta \hat\Fv_\alpha(\rv)
\hat A\rangle \neq \langle \hat\Cv_\alpha(\rv,t) \hat A(t) \rangle$,
as one would expect on general grounds and which is confirmed by the
concrete examples presented below. ii)~Although the sum of the partial
hypercurrent correlation functions is independent of time,
\begin{align}
  & \Sv_A(\rv) + \Cv_A^\rmid(\rv,t)
  +\Cv_A^\rmint(\rv,t) + \Cv_A^\rmext(\rv,t)  = 0,
  \label{EQequilibriumSplitting}
\end{align}
the individual ideal, interparticle, and external contributions each
retain non-trivial temporal dependence; we recall $\Cv_A^\rmid(\rv,t)=
\langle \hat\Cv_\rmid(\rv,t) \hat A(t) \rangle$, {\it etc.} for 'int'
and `ext'.  One might expect this temporal behaviour to occur on
grounds of the general nontrivial setup of these correlation functions
and we exemplify the temporal dependence in specific model situations
in the following.

\begin{figure*}[!t]
  \vspace{1mm}
  \includegraphics[page=1,width=.85\textwidth]{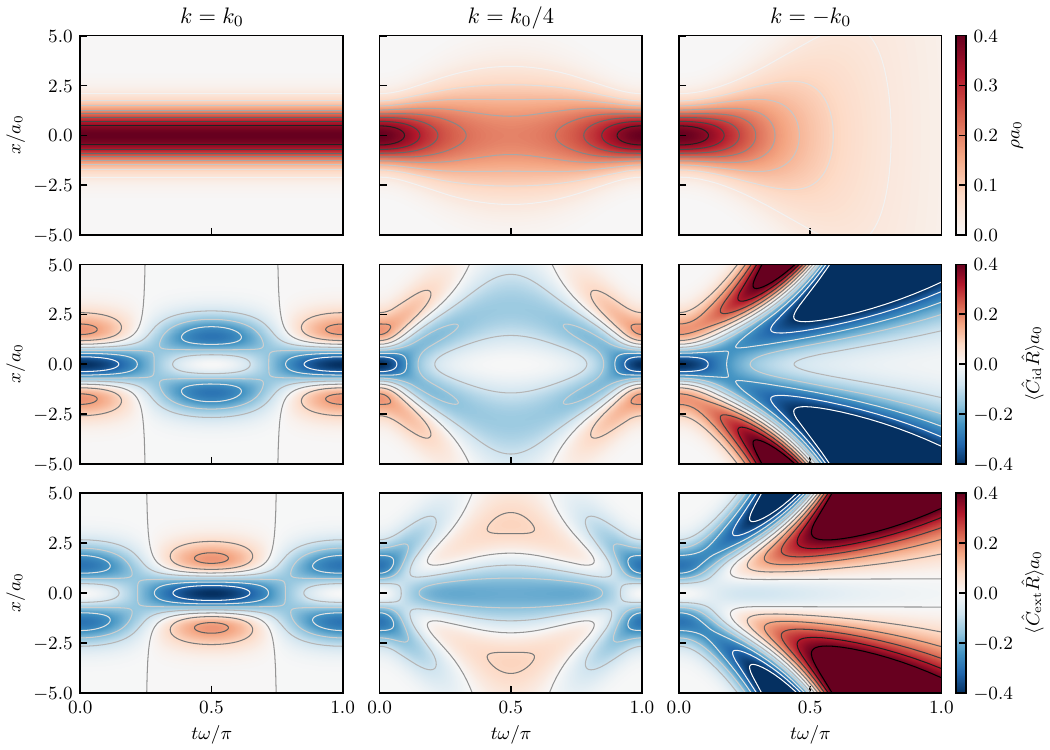}
  \caption{Correlation functions for the ideal gas in a harmonic trap
    in equilibrium (left column) and nonequilibrium (middle and right
    columns). Shown are space-time plots of the scaled dynamical
    density profile $\rho(x,t)\size_0$ (top row) and for the different
    types of hypercurrent correlation functions that arise from
    dynamical gauge invariance, namely the ideal (second row) and
    external (third row) hypercurrent correlation functions for the
    center of mass, $\langle \hat C_\rmid(x,t) \hat R(t)
    \rangle\size_0$ and $\langle\hat C_\rmext(x,t) \hat R(t)
    \rangle\size_0$, with the lengthscale $a_0=1/\sqrt{\beta
      k_0}$. The nonequilibrium sum rule
    \eqr{EQrhoSumRuleOneDimensionDynamic} constrains the sum of these
    three contributions to vanish. Left column: Equilibrium dynamics
    where $H=H_0$ and $k=k_0$ at all times; despite the density
    profile being stationary both hypercurrent correlation functions
    oscillate in time. Middle column: At the initial time the value of
    the spring constant is reduced, $k=k_0/4$, such that temporal
    density oscillations are induced and these affect the hypercurrent
    correlation functions. Right column: At the initial time the sign
    of the harmonic potential is flipped, $k=-k_0$, which creates an
    unstable situation without corresponding equilibrium; pronounced
    structuring is apparent in both hypercurrent correlation
    functions.  }
\label{FIG3}
\end{figure*}

\subsection{Nonequilibrium ideal gas of harmonic oscillators}
\label{SECharmonicOscillators}

As an initial application of the general dynamical gauge invariance
framework to a conrete system, we consider an ideal gas of $N$
particles, with vanishing interparticle potential $u(\rv^N)=0$, in one
spatial dimension.  We demonstrate explicitly that the theory allows
one i)~to recover the correct static hyperforce limit as described in
Sec.~\ref{SEClinkToEquilibrium}, ii) to identify non-trivial time
dependence in equilibrium, iii) to discriminate between thermal
equilibrium states and nonequilibrium stationary states, and vi) to
retain the general theoretical structure when addressing nontrivial
temporal dependence in more general nonequilibrium ideal gas setups.
Furthermore, the analytical solution provides a useful reference for
validation of the simulation methods.

\subsubsection{Setup of the model and dynamical density profile}
\label{SECidealDensityProfile}

To facilitate analytical treatment, we choose the external potential
as being harmonic: $V_\rmext(x)=k x^2/2$, with strength parameter $k$
and one-dimensional position coordinate~$x$. The initial Hamiltonian
has a positive spring constant $k_0>0$, particle mass $m_0$, and
corresponding frequency $\omega_0=\sqrt{k_0/m_0}$. At times $t \geq
0$, we allow in general the constants to be $m\neq m_0$ and $k\neq
k_0$, with corresponding frequency $\omega=\sqrt{k/m}$ and again in
general $\omega\neq\omega_0$.  After the switching, the external
potential can remain confining, $k>0$, or vanish, $k=0$, or become
repulsive, $k<0$.  The latter case constitutes a dynamically unstable
situation with the Hamiltonian being unbounded from below.  The
one-dimensional particle trajectories are $\hat x_i(t)= x_i
\cos(\omega t) + p_i \sin(\omega t)/(m\omega)$ and $\hat p_i(t) = p_i
\cos(\omega t) - m \omega x_i\sin(\omega t)$ with initial state $x_i,
p_i$ at time $t=0$, for all $i=1,\ldots,N$.

Straightforward algebra yields the dynamical density profile
$\rho(x,t)=\langle \hat\rho(x,t)\rangle$ as a normalized Gaussian with
temporally varying width parameter $\alpha$:
\begin{align}
   \rho(x,t) &= N\sqrt{\alpha/\pi}\e^{-\alpha x^2},
   \label{EQrhoAnalyticalOneDimension}   \\
   \alpha &= \beta k \Big/
   \Big[\frac{k}{k_0} + \frac{m_0}{m}
     +\Big(\frac{k}{k_0} - \frac{m_0}{m}\Big)
     \cos(2\omega t) \Big],
   \label{EQalphaAnalytical}
\end{align}
with a scaled version of the expression \eqref{EQalphaAnalytical}
being given in \eqr{EQalphaAnalyticalCompact}.

We summarize several properties of this solution.  Forgoing any
switching and hence retaining $k=k_0 $, $m=m_0$, the system remains in
equilibrium at all times and the width parameter $\alpha_0=\beta
k_0/2=\rm const$, independent of time.  For general switching, $m\neq
m_0$ and $k\neq k_0$, the value of $\alpha$ oscillates in time with
doubled frequency $2\omega$ (we comment on the unstable case at the
end of the section).  The temporal oscillations of $\alpha$ are
bounded by two values, one being $\alpha_0$, which is attained at
times $t=n\pi/\omega$ with $n$ being a nonnegative integer. The
further bounding value of $\alpha$ is $\beta km/(2m_0)$, which is
attained at times $t=(n+1/2)\pi/\omega$. The midpoint value of
$\alpha$ between the two extrema is $\beta k/[(k/k_0)+(m_0/m)]$.

\begin{figure*}[!t]
  \vspace{1mm}
  \includegraphics[page=1,width=.85\textwidth]{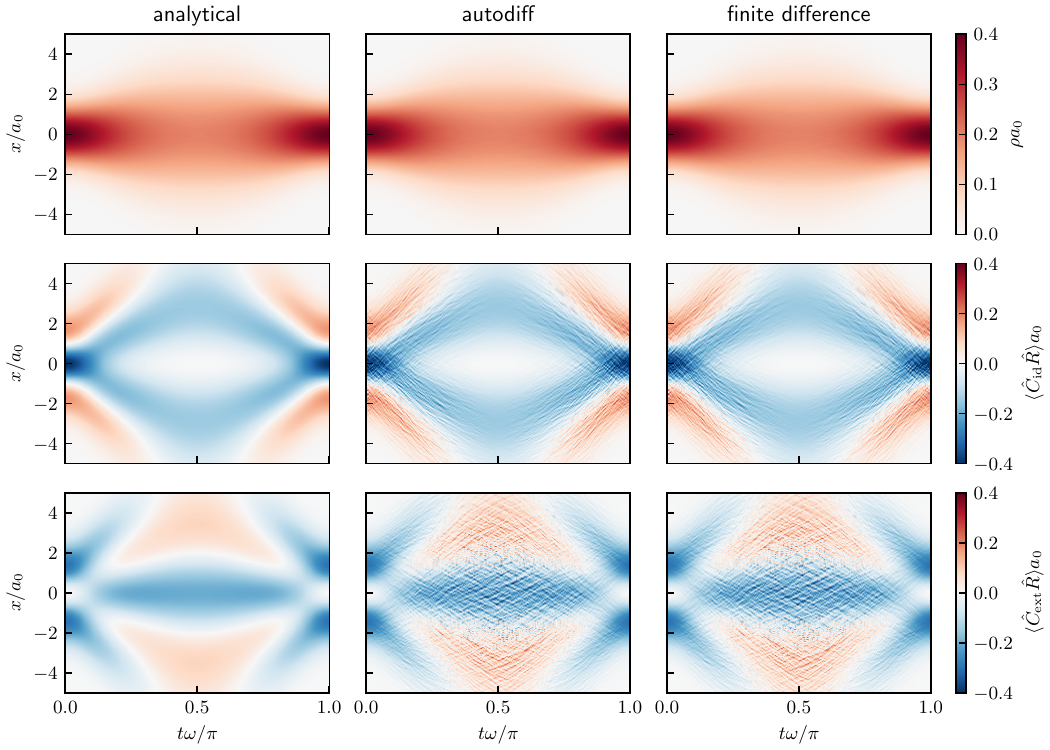}
  \caption{Comparison of results obtained analytically (left column)
    and from molecular dynamics simulations via autodifferentiation
    (center column) and via finite-difference differentiation (right
    column) with respect to the initial state of molecular
    trajectories. Shown is the scaled dynamical density profile
    $\rho(x,t)\size_0$ (top row), the ideal part, $\langle\hat
    C_\rmid(x,t) \hat R(t) \rangle\size_0$, and the external
    contribution, $\langle \hat C_\rmext(x,t) \hat R(t)
    \rangle\size_0$, of the hypercurrent correlation function for the
    scaled center of mass $\hat R$. The results are shown as a
    function of the scaled distance $x/\size_0$, with natural
    lengthscale $\size_0 = 1/\sqrt{\beta k_0}$ and scaled time $t
    \omega/\pi$.  The switching protocol implies $k=k_0/4$ and
    $m=m_0$.  Except for small numerical sampling artifacts, the
    results from all three methods agree with each other to high
    precision and they satisfy the sum rule
    \eqref{EQrhoSumRuleOneDimensionDynamic}.}
\label{FIG4}
\end{figure*}

\begin{figure*}[!t]
  \vspace{1mm}
  \includegraphics[page=1,width=.85\textwidth]{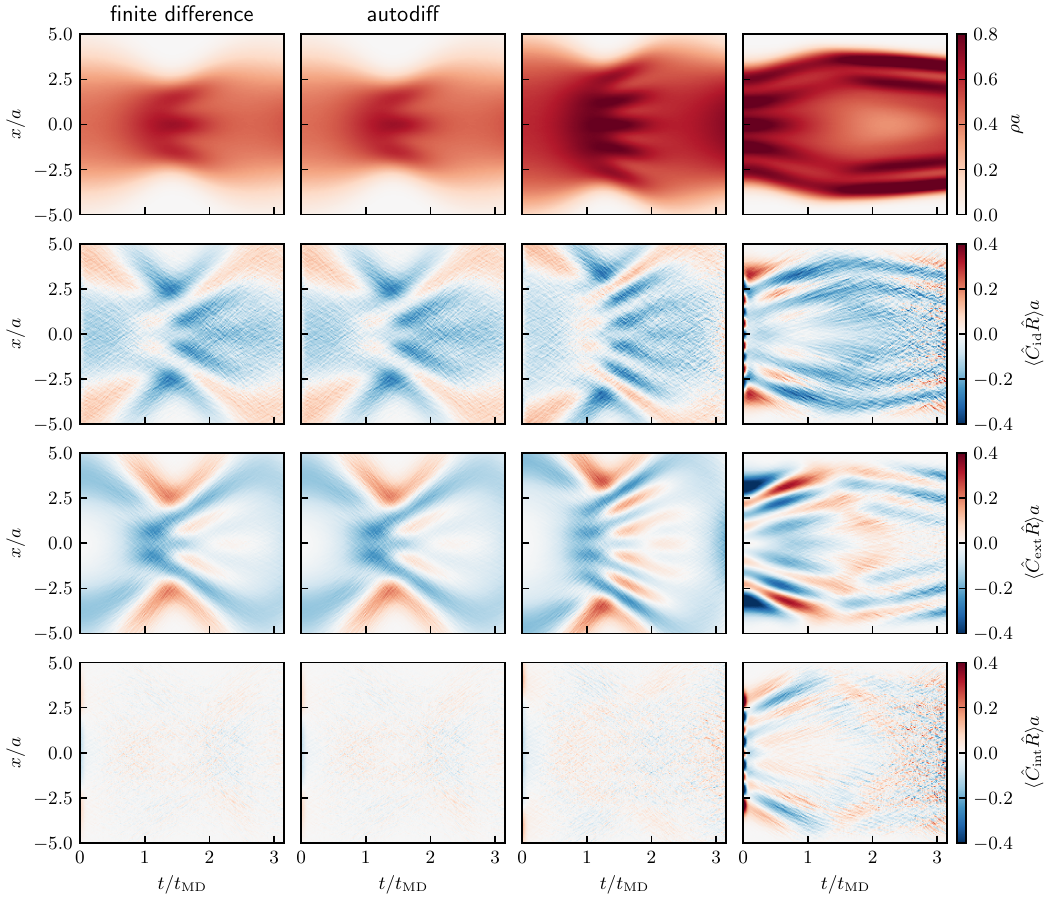}
  \caption{Molecular dynamics results for the hypercurrent correlation
    functions of a one-dimensional system of particles that mutually
    repel via the Weeks-Chandler-Andersen pair potential. Shown is the
    dynamical density profile $\rho(x,t)\size$ (top row) and the three
    contributions to the hypercurrent correlation function for the
    (scaled) center of mass, $\hat A =\hat R$, namely the ideal part
    $\langle \hat C_\rmid(x,t) \hat R(t) \rangle\size$ (second row),
    the external part $\langle \hat C_\rmext(x,t) \hat R(t)
    \rangle\size$ (third row), and the interparticle interaction
    contribution $\langle \hat C_\rmint(x,t) \hat R(t) \rangle\size$
    (bottom row), shown as a function of the scaled distance $x/\size$
    and scaled time $t/\timescale$, where $a$ is the particle size and
    $\timescale=\size\sqrt{m\beta}$ is the time scale. The results are
    shown for a harmonic trap that is switched narrower (first,
    second, and third column) and wider (fourth column), as obtained
    from finite-difference differentiation (first column) and via
    automatic differentiation (second, third, and fourth columns). The
    results from both methods agree with each other for identical
    conditions (first and second columns) and the one-dimensional form
    \eqref{EQcenterOfMassSumRuleOneDimension} of the dynamical sum
    rule \eqref{EQdressedIdentity4} is satisfied with high accuracy.}
\vspace{20mm}
\label{FIG5}
\end{figure*}

As a special case one may choose parameters that satisfy
$km=k_0m_0$. Then despite in general $\omega\neq\omega_0$, the density
profile \eqref{EQrhoAnalyticalOneDimension} remains independent of
time and it thus forms a simple example of a nonequilibrium stationary
state.  The width parameter thereby retains its initial value
$\alpha_0=\beta k_0/2$, unperturbed by this specific switching, over
the course of time.

Figure~\ref{FIG3} displays results for dynamical density profiles
$\rho(x,t)$ (first row) obtained from the analytical solution
\eqref{EQrhoAnalyticalOneDimension} for $m=m_0$ and the three
representative cases of no switching: $k=k_0$, switching to a softer
spring: $k = k_0/4$, and switching to an unstable situation: $k=-k_0$
(from left to right).

\subsubsection{Ideal shift current balance}
\label{SECidealShiftCurrentBalance}

Turning to the gauge theory we first consider the static force density
balance \eqref{EQforceDensityBalanceStatic}, which for the
one-dimensional ideal gas simplifies to:
\begin{align}
  F_{\rmid,0}(x) + F_{\rmext,0}(x) &= 0,
  \label{EQidealForceDensityBalanceStatic}  
\end{align}
where we use scalar notation and the interparticle force density
vanishes, $F_{\rmint, 0}(x)=0$, for the present ideal gas. The static
ideal and external force fields follow from respective explicit
calculation as
\begin{align}
  \frac{\beta F_{\rmid,0}(x)}{\rho_0(x)} &= \beta k_0x,
  \label{EQFidZeroIdealGas}\\
  \frac{\beta F_{\rmext,0}(x)}{\rho_0(x)} &= -\beta k_0x,
  \label{EQFextZeroIdealGas}
\end{align}
where we have normalized by the initial density profile $\rho_0(x)$;
recall that this is a Gaussian defined by
\eqr{EQrhoAnalyticalOneDimension} with $\alpha_0=\beta k_0/2$, such
that $\rho_0(x)=N \sqrt{\beta k_0/(2\pi)}\exp(-\beta k_0x^2/2)$.
The results \eqref{EQFidZeroIdealGas} and \eqref{EQFextZeroIdealGas}
satisfy the static force sum rule
\eqref{EQidealForceDensityBalanceStatic} explicitly.

The dynamic shift current balance \eqref{EQbareHyperForceBalance}
simplifies for the present system to consist of merely two
contributions:
\begin{align}
  C_\rmid(x,t) + C_\rmext(x,t) &= 0,
  \label{EQidealShiftCurrentBalance} 
\end{align}
as the interparticle shift current vanishes, $C_\rmint(x,t)=0$.  Both
the ideal and the external partial shift current is conveniently
normalized by the dynamic density profile $\rho(x,t)$, which leads to
the following odd cubic polynomials in position $x$:
\begin{align}
  \frac{C_\rmid(x,t)}{\rho(x,t)} &= b_1 x + b_3 x^3,
  \label{EQpolynomialCid}
  \\
  \frac{C_\rmext(x,t)}{\rho(x,t)} &= -b_1x - b_3 x^3.
  \label{EQpolynomialCext}
\end{align}
The coefficients $b_1$ and $b_3$ are time-dependent and, when scaled
by the appropriate powers $\alpha$ and $\alpha^2$ of the width
parameter, possess the following forms:
\begin{align}
  &\frac{b_1}{\alpha} =
  \frac{2(\kappa+\kappa_0)\cos(2\omega t)+(\kappa-\kappa_0)
    [5-\cos(4\omega t)]/2}
       {\kappa+\kappa_0 + (\kappa-\kappa_0)\cos(2\omega t)},
       \label{EQpolynomialCoefficientb1}\\
  & \frac{b_3}{\alpha^2} =
  -\frac{(\kappa-\kappa_0)[1-\cos(4\omega t)]}
       {\kappa+\kappa_0+(\kappa-\kappa_0)\cos(2\omega t)}.
       \label{EQpolynomialCoefficientb3}
\end{align}
Here we use the shorthand $\kappa=km$ to denote the product of spring
constant $k$ and mass $m$; correspondingly $\kappa_0 = k_0m_0$ for the
intital state parameters at time $0$.  Using these variables allows
one to express the width parameter \eqref{EQalphaAnalytical}
succinctly as:
\begin{align}
   \alpha &= \frac{\beta k_0\kappa}{
     \kappa + \kappa_0+(\kappa - \kappa_0) \cos(2\omega t) }.
   \label{EQalphaAnalyticalCompact}
\end{align}

It is noteworthy that
Eqs.~\eqref{EQpolynomialCoefficientb1}--\eqref{EQalphaAnalyticalCompact}
share a common denominator; this structure is a consequence of the
scaling with $\rho(x,t)$ in Eqs.~\eqref{EQpolynomialCid} and
\eqref{EQpolynomialCext} and the scaling with powers of $\alpha$ in
Eqs.~\eqref{EQpolynomialCoefficientb1} and
\eqref{EQpolynomialCoefficientb3}.

Considering the limit $t\to 0^+$ leads to the coefficients
\eqref{EQpolynomialCoefficientb1} and
\eqref{EQpolynomialCoefficientb3} attaining respective values
$b_1=\beta k_0$ and $b_3=0$, with the width parameter
\eqref{EQalphaAnalyticalCompact} becoming $\alpha_0 = \beta
k_0/2$. Thus both the ideal and the external contribution to the
dynamic shift current $C(x,0)$ reduce correctly to their corresponding
thermally scaled initial partial force density: $C_\rmid(x,0) = \beta
F_{\rmid,0}(x)$ and $ C_\rmext(x,0) = \beta F_{\rmext,0}(x)$.

We emphasize that the cubic spatial dependence of the (scaled)
external shift current $C_\rmext(x,t)/\rho(x,t)$ given by
\eqr{EQpolynomialCext} is very different from the behaviour of the
time-dependent external force field, which remains linear,
$F_\rmext(x,t)/\rho(x,t)=-kx$, for all times $t\geq 0$.

\begin{figure*}[!t]
  \includegraphics[page=1,width=.53\textwidth]{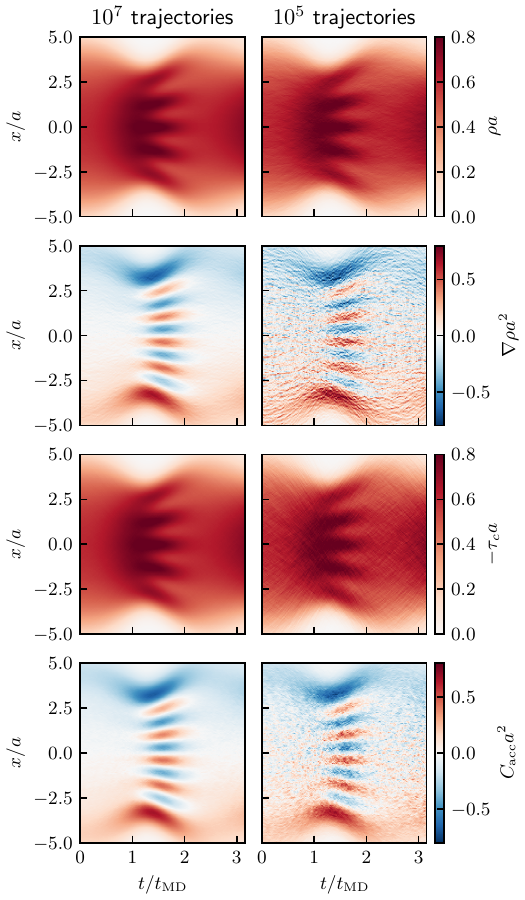}
  \caption{Splitting of the hypercurrent correlation function into
    transport and acceleration contributions. Shown is the scaled
    dynamical density profile $\rho(x,t)\size$ (top row) obtained from
    histogram filling together with its gradient $\nabla \rho(x,t)
    \size^2= \partial\rho(x,t)\size^2/\partial x $ (second row), as
    obtained from numerical differentiation with respect to position
    $x$. The scaled negative mean shift stress $-\tau_C(x,t)\size$
    (third row) is identical, on the scale of the plot, to the density
    profile (first row), which verifies the sum rule
    \eqref{EQtauCsumrule}: $\tau_C(x,t) = - \rho(x,t)$. The scaled
    shift acceleration current $C_\rmacc(x,t)\size$ (bottom row) is
    numerically identical to the density gradient (second row), thus
    verifying the identity \eqref{EQCaccIdentity}:
    $C_\rmacc(x,t)=\nabla\rho(x,t)$, as is demonstrated by
    high-quality averaging over $10^7$ trajectories (left
    column). Reduced averaging over only $10^5$ trajectories (right
    column) reveals reduction of statistical noise of
    $C_\rmacc(x,t)\size$ (right column, bottom panel) as compared to
    the direct result for $\nabla\rho(x,t)a^2$ (right column, second
    panel from top), which renders the shift acceleration current a
    potential candidate for nonequilibrium reduced-variance sampling.}
\vspace{20mm}
\label{FIG6}
\end{figure*}

\subsubsection{Ideal hypercurrent balance}
\label{SECidealHyperCurrentBalance}

As a representative hyperobservable, we consider the sum of positions
$\hat R=\sum_i \hat x_i$; we recall the general hypercurrent sum rule
\eqr{EQdressedIdentity4} for $\hat\Rv(t)$. We first consider the
corresponding static hyperforce balance, which possesses the following
initial state form:
\begin{align}
  & \rho_0(x) + \langle \beta \hat F_{\rmid,0}(x) \hat R\rangle
  +\langle \beta \hat F_{\rmext, 0}(x) \hat R \rangle 
  = 0.
  \label{EQrhoSumRuleOneDimensionStatic}
\end{align}
The two static correlation functions in
\eqr{EQrhoSumRuleOneDimensionStatic} follow from explicit algebra in
the respective forms:
\begin{align}
  \frac{\langle \beta\hat F_{\rmid,0}(x) \hat R \rangle}{\rho_0(x)}
  &= \beta k_0x^2 - 1,
  \label{EQonedFidCorrelator}\\
  \frac{\langle \beta\hat F_{\rmext,0}(x) \hat R \rangle}{\rho_0(x)}
  &= -\beta k_0x^2,
  \label{EQonedFextCorrelator}
\end{align}
where the initial state density profile $\rho_0(x)$ is given below
\eqr{EQFextZeroIdealGas} in its explicit Gaussian form with width
parameter $\alpha_0$.  Again Eqs.~\eqref{EQonedFidCorrelator} and
\eqref{EQonedFextCorrelator} satisfy the static sum rule
\eqref{EQrhoSumRuleOneDimensionStatic}, as can be seen directly by
summing up Eqs.~\eqref{EQonedFidCorrelator} and
\eqref{EQonedFextCorrelator}, adding unity and multiplying by
$\rho_0(x)$.

The hypercurrent sum rule \eqref{EQdressedIdentity4} for the
one-dimensional ideal gas attains the form:
\begin{align}
  \rho(x,t) + \langle \hat C_\rmid(x,t) \hat R(t) \rangle 
  + \langle \hat C_\rmext(x,t) \hat R(t) \rangle 
  & = 0,
  \label{EQrhoSumRuleOneDimensionDynamic}
\end{align}
where $\hat C(x,t)=\hat C_\rmid(x,t) + \hat C_\rmext(x,t)$ is the
one-dimensional version of the shift current operator with $\hat
C_\rmint(x,t)=0$, as the particles are ideal.
The two hypercurrent correlation functions in
\eqr{EQrhoSumRuleOneDimensionDynamic} follow by explicit calculation
as two even fourth-order polynomials in~$x$, when normalized by the
dynamic density profile:
\begin{align}
  \frac{\langle \hat C_\rmid(x,t) \hat R(t) \rangle}{\rho(x,t)} &=
  -1 + b_0  + b_2 x^2 + b_4 x^4,
  \label{EQonedCidCorrelator}\\
  \frac{\langle \hat C_\rmext(x,t) \hat R(t) \rangle}{\rho(x,t)} &=
  -b_0 - b_2 x^2 - b_4 x^4.
  \label{EQonedCextCorrelator}
\end{align}
Again, the solution \eqref{EQonedCidCorrelator} and
\eqref{EQonedCextCorrelator} satisfies the dynamical sum rule
\eqref{EQrhoSumRuleOneDimensionDynamic}, as follows straightforwardly.

The coefficients $b_0$, $b_2$, and $b_4$ in
Eqs.~\eqref{EQonedCidCorrelator} and \eqref{EQonedCextCorrelator} are
time-dependent in general. Scaling by the respective appropriate
powers $1,\alpha$, and $\alpha^2$ of the time-dependent Gaussian width
parameter $\alpha$, see \eqr{EQalphaAnalyticalCompact} for its
explicit form, yields:
\begin{align}
  b_0 &= 
  \frac{\kappa_0 [1-\cos(2\omega t)]}
  {\kappa+\kappa_0+(\kappa-\kappa_0)\cos(2\omega t)},
  \label{EQcoefficient0}  \\
  \frac{b_2}{\alpha} &= 
  \frac{2(\kappa+\kappa_0)\cos(2\omega t)+
  (\kappa-\kappa_0)[3-\cos(4\omega t)]}
  {\kappa+\kappa_0+(\kappa-\kappa_0)\cos(2\omega t)},
  \label{EQcoefficient2}\\
  \frac{b_4}{\alpha^2} &= -\frac{(\kappa-\kappa_0)[1-\cos(4\omega t)]}
   {\kappa+\kappa_0+(\kappa-\kappa_0)\cos(2\omega t)},
  \label{EQcoefficient4}
\end{align}
where as before $\omega=\sqrt{k/m}$ is the oscillator frequency after
switching and the commonality of denominators, as observed in
Eqs.~\eqref{EQpolynomialCoefficientb1}--\eqref{EQalphaAnalyticalCompact},
is retained. Comparison of \eqr{EQcoefficient4} to the coefficient
\eqref{EQpolynomialCoefficientb3} for the shift current reveals that
$b_3=b_4$.

As a consistency check of the static and dynamical solutions, at time
$t=0$ and hence $\alpha_0 = \beta k_0/2$, the dynamical coefficients
\eqref{EQcoefficient0}--\eqref{EQcoefficient4} reduce to: $b_0=0,
b_2=\beta k_0$, and $b_4=0$. Inserting these results into the
hypercurrent correlation functions \eqref{EQonedCidCorrelator} and
\eqref{EQonedCextCorrelator} reduces these expressions correctly to
the respective static hyperforce correlation functions
\eqref{EQonedFidCorrelator} and \eqref{EQonedFextCorrelator}, such
that indeed $\langle \hat C_\rmid(x,0) \hat R(0) \rangle = \langle
\beta \hat F_{\rmid,0}(x) \hat R \rangle$ and $\langle \hat
C_\rmext(x,0) \hat R(0) \rangle = \langle \beta \hat F_{\rmext,0}(x)
\hat R \rangle$.

The ideal and external parts of the hypercurrent correlation function,
$\langle \hat C_\rmid(x,t)\hat R(t)\rangle$ and $\langle \hat
C_\rmext(x,t)\hat R(t)\rangle$, as respectively given by
Eqs.~\eqref{EQonedCidCorrelator} and \eqref{EQonedCextCorrelator}, are
depicted graphically in the second and third row of Fig.~\ref{FIG3}.
We recall the above description of the three considered cases of no
switching (first column), switching to softer confinement (second
column), and switching to an unstable situtation by reversing the sign
of the force constant (third column); we return to the latter case at
the end of Sec.~\ref{SECidealLimits}. As anticipated in the discussion
given in Sec.~\ref{SEClinkToEquilibrium} on the basis of general
arguments, the hyperforce correlation funtions indeed display
nontrivial time dependence, and they provide arguably much deeper
insight into the dynamical structuring than does the dynamical density
profile (first row in Fig.~\ref{FIG3}).

\subsubsection{Reduction to limiting cases of the ideal gas}
\label{SECidealLimits}

As a specific simple situation, it is interesting to consider the
dynamics when the model parameters are kept constant at the initial
time, $k=k_0$ and $m=m_0$, and hence no switching occurs. Then $b_1 =
2 \alpha_0 \cos(2\omega_0 t) = \beta k_0 \cos(2\omega_0 t)$ from
\eqr{EQpolynomialCoefficientb1} and $b_3 = 0$ from
\eqr{EQpolynomialCoefficientb3}. Hence the ideal and external mean
shift current both follow from Eqs.~\eqref{EQpolynomialCid} and
\eqref{EQpolynomialCext} as $C_\rmid(x,t) = -C_\rmext(x,t) = \beta k_0
x \cos(2\omega_0 t) \rho_0(x)$, where we recall $\alpha_0 = \beta
k_0/2$ as the initial width parameter and $\omega_0 = \sqrt{k_0/m_0}$
as the initial frequency and note the period doubling effect.
Furthermore, for the present case of switching being absent, the
hypercurrent identity \eqref{EQrhoSumRuleOneDimensionDynamic}, reduces
from the general nonequilibrium solution
\eqref{EQcoefficient0}--\eqref{EQcoefficient4}
similarly to $b_0 = [1-\cos(2\omega_0 t)]/2,$ $ b_2 = \beta
k_0\cos(2\omega_0 t)$, and $b_4 = 0$, which simplifies the ideal
\eqref{EQonedCidCorrelator} and external \eqref{EQonedCextCorrelator}
hypercurrent correlation functions.  Hence the occurring oscillation
is characterized by a doubled initial frequency $2\omega_0$. It is
remarkable that even in this arguably simplest setup of noninteracting
harmonic oscillators in thermal equilibrium, there is nontrivial
temporal dependence exposed by the hypercurrent approach; we recall
the illustration of the ideal and external hypercurrent correlation
functions in the first column of Fig.~\ref{FIG3}.

Turning to the case of nonequilibrium steady states with constrained
parameter choices $km=k_0m_0$, the coefficients of the shift current
are $b_1 = \beta k_0 \cos(2\omega t)$, $b_3=0$ and hence $C_\rmid(x,t)
= -C_\rmext(x,t) = \beta k_0 x \cos(2\omega t) \rho_0(x)$, where the
frequency is twice that {\it after} switching: recall $\omega
=\sqrt{k/m}$ and that this is in general different from the doubled
frequency $2\omega_0$ identified above in equilibrium. We recall that
the density profile itself remains stationary in the present case. The
hypercurrent coefficients are $b_0=[1-\cos(2\omega t)]/2 $,
$b_2=2\alpha_0 \cos(2\omega t) = \beta k_0 \cos(2\omega t)$, and
$b_4=0$. 

As a final case, the presented general solution for the contributions
to the shift current identity (Sec.~\ref{SECidealShiftCurrentBalance})
and to the hypercurrent balance
(Sec.~\ref{SECidealHyperCurrentBalance}) remains valid when the
external potential is no longer confining and thus $k<0$. Then the
trigonometric dependence becomes hyperbolic, such that e.g.,
$\cos(2\omega t) = \cosh(2|\omega| t)$, where
$|\omega|=\omega/i=\sqrt{-k/m} \geq 0$ with imaginary unit $i$, which
again results in non-trivial time dependence in the dynamical
correlation functions \eqref{EQonedCidCorrelator} and
\eqref{EQonedCextCorrelator}, see the last column of Fig.~\ref{FIG3}
for the graphical representations.
Taking the limit $k\to 0$, such that the system is free at $t>0$,
allows one to make further simplifications to the above analytical
results.

\begin{figure*}[!t]
  \includegraphics[page=1,width=.49\textwidth]{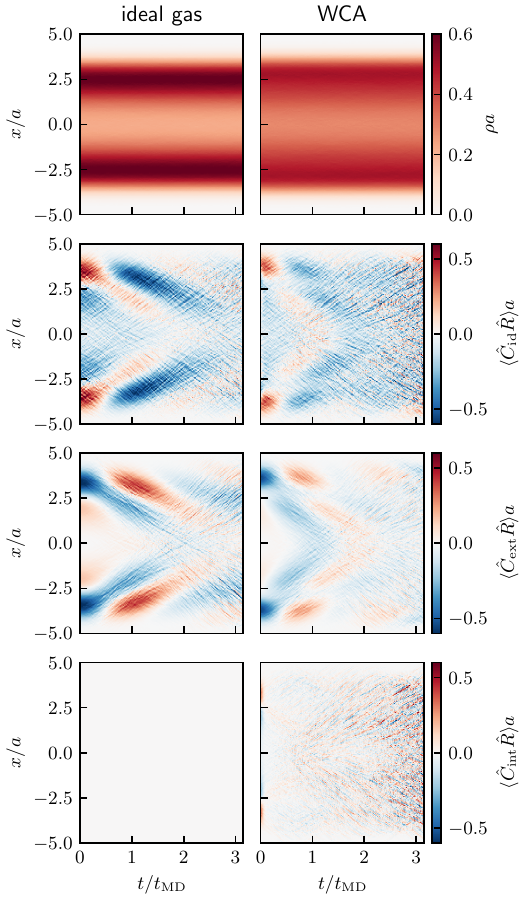}
  \caption{Dynamical gauge correlation functions of the ideal gas
    (left column) and of mutually interacting particles (right column)
    inside of an external double-well potential. Shown is the scaled
    dynamical density profile $\rho(x,t)\size$ (top row) and the
    hypercurrent correlation function for the total position,
    consisting of the ideal part, $\langle C_\rmid(x,t) R(t)\rangle
    \size$ (second row), external part, $\langle C_\rmext(x,t)
    R(t)\rangle \size$ (third row), and interparticle contribution,
    $\langle C_\rmint(x,t) R(t)\rangle \size$ (bottom row).  The
    system is in equilibrium and thus the density profile is
    stationary. Note the spatiotemporal hypercurrent structure that is
    indicative of barrier crossing.}
\label{FIG7}
\end{figure*}

\subsection{Molecular dynamics simulations}
\label{SECmolecularDynamicsResults}

\subsubsection{Simulation methodology}
\label{SECmdValidation}

We turn to simulations to gain further and deeper physical insight
into the nature of the hypercurrent correlation structure beyond the
simple nonequilibrium ideal gas setup of
Sec.~\ref{SECharmonicOscillators}. We base our methodology on the
trajectory level, as described in Sec.~\ref{SECtrajectories} and in
particular on the initial state differentiation laid out in
Sec.~\ref{SECmolecularDynamicsConcepts}.

We use two different methods to realize the initial state time
differentiation. As a basis for the time evolution we use in both
methods the velocity Verlet algorithm \cite{verlet1967, hansen2013,
  frenkel2023book}, see Sec.~\ref{SECtrajectories}, but we see no
reason why one should not be able to use other molecular dynamics time
integrators \cite{frenkel2023book}. First, to realize the initial
state derivative, we use automatic differentiation to keep track of
the effect of differential changes in the initial microstate of each
molecular dynamics trajectory. Results for the shift current and for
the hypercurrent correlation functions are then obtained according to
the trajectory-based picture described in
Sec.~\ref{SECmolecularDynamicsConcepts}. Implementation-wise, the
method requires the molecular dynamics to be performed in an
environment that provides ready access to the automatic
differentiation functionality; we use a custom molecular dynamics
implementation in the programming language Julia
\cite{mueller2024dynamicRepository} and note recent progress in
differentiable molecular dynamics \cite{schoenholz2020}, where
libraries are available.

As a second and alternative route, we perform differentiation via
finite differences, which can be realized in any molecular dynamics
environment. Here the initial state differentiation is based on
dynamically propagating one additional trajectory that differs from
the corresponding original trajectory by a single, additional time
step that is performed on the initial microstate and with respect to
the dynamics generated by the initial state Hamiltonian~$H_0$. The
magnitude of this time step can be chosen freely, as described below.
The thus altered initial microstate is then propagated forward in
time, in an identical way as the original corresponding trajectory,
i.e., on the basis of the given form of $H$ for $t\geq 0$.  At (each)
target time~$t$, the finite difference with the unperturbed trajectory
then provides the numerator for the finite difference ratio; the
denominator is the time step used for the perturbation of the initial
state. To perform the dynamical evolution, we use a standard value of
the size of the time step, $\Delta t=5\times 10^{-3}\timescale$, with
microscopic time scale $\timescale=\size\sqrt{m\beta}$, where $a$ is
the particle size (as specified below). The initial time derivative is
performed with a single, much smaller initial time step of
$10^{-4}\timescale$.

To first validate both simulation methods we use the harmonic
oscillator setup, as described above in
Sec.~\ref{SECharmonicOscillators}, and compare against the analytical
solution. The use of the simulation methods in this case also serves
to demonstrate and illustrate the arguably more intuitive access that
the trajectory-based picture provides for the dynamical gauge
theory. Despite the simplicity of the noninteracting sytem, we deem
the test to be nontrivial, due to the significant conceptual
differences between the three methods (analytical solution, automatic
and finite-difference trajectory differentiation) to obtain the
respective results. We recall the explicit phase space manipulations,
described in Sec.~\ref{SECharmonicOscillators}, to obtain the
analytical solution and the trajectory-based formulation of the
initial state time derivative in
Sec.~\ref{SECmolecularDynamicsConcepts}.

In Fig.~\ref{FIG4} we use the analytical solution as a reference (left
column) to compare the results from both simulation methods: automatic
differentiation (center column) and finite-difference differentiation
(right column). As expected, the results for the dynamical density
profile $\rho(x,t)$ are numerically identical. We recall that based on
the simulation, mere filling of a position- and time-resolved
histogram of particle positions is required. Averaging over the initial
ensemble, as is realized via Monte Carlo simulations on the basis of
the initial state Hamiltonian $H_0$ and initial inverse temperature
$\beta$, then yields the numerical results. 

Obtaining simulation results for the ideal and external hypercurrent
correlation functions, $\langle \hat C_\rmid(x,t) \hat R(t) \rangle$
and $\langle \hat C_\rmext(x,t) \hat R(t) \rangle$, requires to carry
out explicitly the initial state time differentiation, as is apparent
from the definition \eqref{EQdefinitionOfBare} of the shift current
observable.  Figure~\ref{FIG4} demonstrates excellent agreement of the
results from both simulation methods with each other and with the
analytical solution. Besides a strong validation of the computational
methodology (we compare both simulation methods against each other for
mutually interacting particles below) this also confirms the
successful mirroring of the dynamical gauge invariance theory, as
developed in Sec.~\ref{SECdynamicalGaugeInvariance} on the basis of
phase space differential operator methods, using the arguably more
intuitive trajectory pricture; see the description in
Secs.~\ref{SECtrajectories} and \ref{SECmolecularDynamicsConcepts}.

\subsubsection{Interacting particles in harmonic confinement}
\label{SECmolecularDynamicsWithInteractions}

Both molecular dynamics simulation methods allow one to go beyond
noninteracting systems in a relatively straightforward way, as the
required time evolution remains based on temporal discretization, here
via the velocity Verlet algorithm, and the initial state time
derivative remains identical. As a crucial difference to the ideal
gas, the interparticle shift current observable $\hat C_\rmint(x,t)$
no longer vanishes, and all further shift and hypercurrent observables
are naturally affected by the altered dynamics.

We consider mutually repulsive particles that interact with a pair
potential $\phi(r)$ as a function of (one-dimenional) interparticle
distance $r$.  Figure~\ref{FIG5} displays results for the
Weeks-Chandler-Andersen pair potential as a prototypical short-ranged,
strongly repulsive model. Explicitly the pair potential is thereby
given by $\phi(r) = 4\epsilon[(r/\size)^{-12}- (r/\size)^{-6}+1/4]$
for $r \leq 2^{1/6}\size$ and zero otherwise, where $\size$ is a
lengthscale, $\epsilon$ is an energy scale, and we choose temperature
such that $\beta\epsilon=1$.

The general center-of-mass sum rule \eqref{EQdressedIdentity4} attains
the following one-dimensional split form:
\begin{align}
  &  \rho(x,t) + \langle \hat C_\rmid(x,t) \hat R(t) \rangle
 + \langle \hat C_\rmint(x,t) \hat R(t) \rangle
 \notag\\&\quad\quad\;\;\,
 + \langle \hat C_\rmext(x,t) \hat R(t) \rangle = 0.
 \label{EQcenterOfMassSumRuleOneDimension}
\end{align}
Figure~\ref{FIG5} displays results for the density profile and the
further three hypercurrent correlation functions on the left hand side
of \eqr{EQcenterOfMassSumRuleOneDimension} using various different
switching setups; see the caption of Fig.~\ref{FIG5}. The simulation
results from automatic differentiation agree with those from the
finite-difference method within high numerical accuracy. Besides
providing a consistency check, this agreement validates each method as
being fit for applying the hypercurrent framework to systems of
mutually interacting particles. Despite our test system being only
one-dimensional, the fact that both differentiation methods perform
well can also serve as a vindication of the implied concept of initial
state differentiation, as it arises from the dynamical gauge
invariance.

\subsubsection{Towards reduced-variance sampling}
\label{SECaccelerationTransportValidation}

We have thus far presented simulation results based on the splitting
of the hypercurrent correlation function into ideal, interparticle,
and external contributions as they arise from the splitting of the
initial state Liouvillian $L_0$; we recall \eqr{EQLsplitting}. We next
address the splitting of the hypercurrent observable into acceleration
and transport contributions presented in
Sec.~\ref{SECmolecularDynamicsConcepts}; see specifically
\eqr{EQCvSplittingAccelerationTransport} therein. For the present
situation, we express the general relationships \eqref{EQtauCsumrule}
and \eqref{EQCaccIdentity} in the following one-dimensional forms:
\begin{align}
  \tau_C(x,t) &= -\rho(x,t),
  \label{EQtauCsumruleOneDimension}\\
  C_\rmacc(x,t) &= \nabla \rho(x,t),
  \label{EQCaccIdentityOneDimension}
\end{align}
where $\nabla=\partial/\partial x$ and the general $d$-dimensional
shift current sum rule \eqref{EQbareHyperForceBalanceSplitForm}
simplifies to $C_\rmacc(x,t)+\nabla \tau_C(x,t)=0$.

We consider the case of switching to a stiffer trap and present the
corresponding dynamical density profile $\rho(x,t)$ together with its
numerically differentiated gradient, $\nabla\rho(x,t)$, in the first
and second rows of Fig.~\ref{FIG6}. Thereby the results for the
density profile are obtained from histogram-resolved counting of
particle positions. Then calculating the numerical position derivative
of the data gives results for the density gradient, which displays the
typical effect of amplification of the statistical noise.

The third panel in Fig.~\ref{FIG6} shows results for the mean shift
stress tensor \eqref{EQtauChatDefinition}, which we spell out in one
dimension as: $\tau_C(x,t) = -\beta \langle \sum_i \delta(x-\hat
x_i(t))\hat x_i^\bullet(t)\hat p_i(t)\rangle$; we recall that the bold
dot denotes the initial-time derivative \eqref{EQAhatBullet}. The
simulation results for $\tau_C(x,t)$ are numerically nearly identical
to those in the first row of Fig.~\ref{FIG6}, which validates the sum
rule \eqref{EQtauCsumruleOneDimension}.
The general definition of the acceleration part
\eqref{EQCacchatDefinition} of the shift current, as is relevant for
\eqr{EQCaccIdentityOneDimension}, attains the form:
$C_\rmacc(x,t)=\beta\langle \sum_i \delta(x-\hat x_i(t)) \hat
p_i^\bullet(t)\rangle$. The corresponding simulation results are shown
in the fourth row of Fig.~\ref{FIG6}. It is noteworthy that these
results i) are consistent with those presented in the second row for
$\nabla\rho(x,t)$, as expected from the sum rule
\eqref{EQCaccIdentityOneDimension}, and ii) that they show reduced
statistical noise -- see the right column of Fig.~\ref{FIG6} where the
number of trajectories has been reduced to demonstrate the effect.

Hence, strikingly, each of the sum rules
\eqref{EQtauCsumruleOneDimension} and
\eqref{EQCaccIdentityOneDimension} provides an independent means of
{\it practical} access to the dynamical density profile
$\rho(x,t)$. (Using \eqr{EQCaccIdentityOneDimension} requires spatial
integration to undo the gradient operation; see
e.g.\ Refs.~\cite{rotenberg2020, delasheras2018forceSampling,
  coles2021} that address this point in equilibrium.) When using the
finite-difference method to carry out the initial time derivative, the
computational overhead involves a mere factor 2. The thus obtained
results display reduced statistical variance over the counting
method. Hence the present methodology offers a way forward to
generalize equilibrium reduced-variance (force-sampling) methods
\cite{rotenberg2020, delasheras2018forceSampling, coles2021} to {\it
  nonequilibrium} situations. This could potentially be very general,
when starting with the hypercurrent identities \eqref{EQJohanna2} and
\eqref{EQJohanna1} that hold for general observables $\hat A$. We have
verified numerically both of these identities in the present system
and for the present choice of hyperobservable (results not shown).

\subsubsection{Double-well potential and barrier crossing}
\label{SECdoubleWell}

To demonstrate that genuine physical insight can be gained from the
present framework, we turn to the classical double-well problem and
show in Fig.~\ref{FIG7} results for both the ideal gas and for
repulsive Weeks-Chandler-Andersen particles. As a choice for a
specific situation, we keep the external double-well potential
unchanged over time, $V_\rmext(x) = \epsilon (x^2 - x_d^2)^2 / x_d^4$,
with constant distance $x_d = 2.5 a$ between the maximum and each
minimum; $\epsilon$ is the energy scale of the pair interaction
potential and $a$ is the particle size and we consider
$\beta\epsilon=1$ as before.  The density profile remains stationary
in this situation and it is identical to its initial equilibrium form
(see the panels in the top row of Fig.~\ref{FIG7}). Both the ideal and
external parts of the hypercurrent correlation function exhibit
pronounced spatiotemporal structure; we recall the sum rule
\eqref{EQdressedIdentity4}, which we find again to be satified with
high numerical accuracy.

The spatiotemporal structuring of the hypercurrent correlation
functions shows clear signs of barrier crossings, see the diagonal
streaks in Figs.~\ref{FIG7}. Comparing the behaviour of the
interacting system with that of the ideal gas reveals that the
presence of the repulsive interparticle interactions leads to more
immediate barrier crossing, as is arguably consistent with
intuition. We conclude that the present setup offers fresh insight
into the coupled motion in complex energy landscapes and leave more
detailed investigations to future work.

\section{Conclusions}
\label{SECconclusions}

In conclusion, we have explored the consequences of dynamical gauge
invariance against phase space shifting in the nonequilibrium
statistical mechanics of many-body systems. The shifting
transformation has been identified previously for systems in thermal
equilibrium \cite{mueller2024gauge, mueller2024whygauge}, where it was
shown to induce exact static identities, ranging from global force
\cite{hermann2021noether} and variance identities
\cite{hermann2022variance}, force-force correlation `3g'-sum rules
\cite{sammueller2023whatIsLiquid}, to locally resolved quantum
\cite{hermann2022quantum} and classical \cite{tschopp2022forceDFT}
force balance relationships. The framework also allows one to obtain
very general hyperforce correlation identities
\cite{robitschko2024any} which are embedded, via an associated
generalized ensemble, in hyperdensity functional theory for the
behaviour of general observables in equilibrium
\cite{sammueller2024hyperDFT, sammueller2024whyhyperDFT}.

The derivation of these prior equilibrium results was based primarily
on variational calculus and in particular on exploiting the properties
of functional derivatives with respect to the shifting field that
parameterizes the transform on phase space. As the shifting field
itself plays the role of a mere gauge function \cite{mueller2024gauge,
  mueller2024whygauge}, here we have rather worked with the equivalent
differential operator formalism~\cite{mueller2024gauge,
  mueller2024whygauge}.  The higher level of abstraction that the
operator method provides over the variational method leads to
significant simplification of the complexity of the required algebra.
We have shown that the static shifting operators $\bsig(\rv)$ acquire
time dependence via the standard embedding inside of a propagator
`sandwich' as given by the time evolution
\eqref{EQsigmaDynamicalAsPropagatorSandwich} for $\bsig(\rv,t)$. This
particular propagator structure might be more familiar from quantum
theory \cite{sakurai1973book} than it is within the present classical
physics, but it indeed also constitutes a general property of
(classical) differential operators, as laid out in
Sec.~\ref{SECtimeEvolution}.

The mechanism of temporal nonlocality that is inherent in the
dynamical gauge invariance is markedly different from the more common
memory integral formalism, as is central in modelling via generalized
Langevin equations \cite{schilling2022, zwanzig2001, hansen2013} and
power functional theory \cite{schmidt2022rmp, treffenstaedt2020shear,
  treffenstaedt2021dtpl, renner2022prl}.  Rather than a temporal
integral, the gauge framework features a time differential structure,
which arises from a specific differential change according to the time
evolution of the initial equilibrium ensemble. While this operation
can be seen technically as a perturbation, it is also arguably the
most natural one, as it is inherent in the initial ensemble and the
initial thermal distribution function remains invariant under the
change; we recall the illustration shown in Fig.~\ref{FIG1}.

Our primary general results are the shift current balance
\eqref{EQshiftingCurrentVanishes} and the hypercurrent identity
\eqref{EQdressedHyperForceBalance}, which are both exact. These sum
rules acquire via decomposing into ideal (kinetic), interparticle, and
external contributions the respective
forms~\eqref{EQbareHyperForceBalance} and
\eqref{EQdressedHyperForceBalanceSplitForm}. We have considered four
concrete examples of hyperobservables, where in particular the
correlation function of the sum of all positions with the shift
current equals the (negative) dynamical density profile, see
\eqr{EQdressedIdentity4}.

The harmonically confined ideal gas served as a toy model to
illustrate some of the properties of the shift and hypercurrent
correlation function. An initial thermal ensemble is thereby set into
motion by switching both the spring constant and the particle mass in
the most general case. Given the simplicity of the setup, the
resulting rich gauge correlation structure is striking, as it
displays, e.g., period doubling effects and polynomial contributions
in position of higher order than one might expect naively to find in a
system of harmonic oscillators.

Our simulation results have confirmed the above picture via
reproducing the analytical ideal gas solution, which provides a
consistency check, and via enabling us to address the effects of
interparticle interactions on the gauge correlation behaviour.  It is
noteworthy that the effects of interparticle interaction are
incorporated effortlessly into the setup. These effects are contained
in an interparticle contribution to the shift current, which upon
averaging forms the mean interparticle shift current and furthermore
consitutes the correlator contribution in the hypercurrent sum rule
\eqref{EQdressedHyperForceBalanceSplitForm}.  All contributions to the
shift current observable $\hat \Cv(\rv,t)$ are thereby accessible via
automatic differentiation; we recall our description of the virtues of
the method in the introduction
(Sec.~\ref{SECintroduction}). Nevertheless, working with finite
differences is entirely feasible, as we have demonstrated; see the
comparison of both simulation methods presented in Fig.~\ref{FIG4} and
in Fig.~\ref{FIG5} (see its two leftmost columns).  The
finite-difference method requires only very moderate overhead over
standard molecular dynamics work, as one merely needs to analyze the
differences of pairs of trajectories that differ by a small finite
change in their initial microstates.

It is worth pointing out several differences of the dynamical gauge
invariance theory with several established methods in nonequilibrium
statistical mechanics.  The present framework is exact and no
approximations are involved in the treatment of the many-body physics
as set up in Sec.~\ref{SECstatisticalMechanics}. Yet, investigating
the combination with approximate closure relations could be
worthwhile.  The central concept of initial state time
differentiation, as described in
Sec.~\ref{SECinitialStateTimeDifferentiation}, is different from
standard dynamical perturbation analysis against generic changes in
the initial conditions. Here the change in initial conditions is
generated by the application of the initial state Liouvillian $L_0$,
which is natural, as the initial thermal ensemble is invariant under
its action to generate the time evolution of the initial state.

The dynamical gauge invariance leads to shift current and hypercurrent
sum rules that are structurally different from the Jarzynski equation
\cite{jarzynski1997} and the further fluctuation theorems of
stochastic thermodynamics \cite{seifert2012}. Yet, exploring
connections and possible cross-fertilization with stochastic
thermodynamics constitutes an interesting topic for future work.
While the present treatment is based genuinely on dynamical averages,
as is standard procedure \cite{zwanzig2001}, the dynamical gauge
invariance framework is distinct from the projection operator
formalism and from mode-coupling theory. Investigating the
consequences for these approaches is a valuable point for future work.
In particular tracing out the connections with the Yvon theorem
\cite{yvon1935, hansen2013}, as used in G\"otze's seminal account of
mode-coupling theory \cite{goetze2009}, constitutes an intereresting
topic.

The mapped averaging framework developed by Kofke and coworkers
\cite{moustafa2015, schultz2016, moustafa2017jctp, moustafa2017prb,
  schultz2018, purohit2018, moustafa2019, purohit2020, moustafa2022,
  lin2018, trokhymchuk2019, schultz2019, purohit2019} is a highly
efficient sampling scheme for equilibrium properties of complex
systems. Whether the present methodology could help to generalize the
mapped averaging to nonequilibrium situations is an interesting
question.  Furthermore, possible generalizations of force sampling
\cite{borgis2013, delasheras2018forceSampling,
  renner2023torqueSampling, rotenberg2020, purohit2019, coles2019,
  coles2021, mangaud2020, coles2023revelsMD}, as used previously for
local transport coefficients and mobility profiles within the
Green-Kubo formalism \cite{rotenberg2020, mangaud2020}, could be very
interesting. The reduced-variance effect described in
Sec.~\ref{SECaccelerationTransportValidation} points to the practical
feasibility.  Finally, the connection to power functional theory
\cite{schmidt2022rmp, schmidt2018md} is worth exploring and whether
the present formalism can shed further light on the limitations of the
dynamical density functional theory \cite{delasheras2023perspective}.
Potential applications of the dynamical gauge sum rules include the
development of convergence tests and sampling schemes for simulations
and to provide consistency checks for dynamical neural functionals
\cite{delasheras2023perspective, zimmermann2024ml, schmidt2022rmp}.

\bigskip

\begin{acknowledgments}
We acknowledge useful discussions with Sophie Hermann and Thomas
Kriecherbauer.  This work is supported by the DFG (Deutsche
Forschungsgemeinschaft) under project no.~551294732.
\end{acknowledgments}

\begin{center}{\bf DATA AVAILABILITY}\end{center}

The data that support the findings of this study are openly available
\cite{mueller2024dynamicRepository}.


\end{document}